\documentclass[prd,preprint,nofootinbib]{revtex4-1}
\pdfoutput=1

\usepackage{amssymb,amsmath}
\usepackage{color}
\usepackage{graphicx,subfigure}
\usepackage{hyperref}
\usepackage{float} 
\usepackage[T1]{fontenc}
\usepackage{bbm}

\newcommand{\nn}{\nonumber}

\newcommand \vect[1]{
        \left(\begin{matrix} #1 \end{matrix}\right)
}

\newcommand \ket[1]{
        \left| #1 \right>
}

\newcommand \bra[1]{
        \left< #1 \right|
}

\begin{document}

\title{SU(3)$_F$ Analysis for Beauty Baryon Decays} 

\author{Avital Dery}
\email{avital.dery@cornell.edu}
\affiliation{Department of Physics, LEPP, Cornell University, Ithaca, NY 14853, USA}
\author{Mitrajyoti Ghosh}
\email{mg2338@cornell.edu}
\affiliation{Department of Physics, LEPP, Cornell University, Ithaca, NY 14853, USA}
\author{Yuval Grossman}
\email{yg73@cornell.edu}
\affiliation{Department of Physics, LEPP, Cornell University, Ithaca, NY 14853, USA}
\author{Stefan Schacht}
\email{ss3843@cornell.edu}
\affiliation{Department of Physics, LEPP, Cornell University, Ithaca, NY 14853, USA}

\begin{abstract}
We perform a general SU(3)$_F$ analysis of $b\rightarrow c\bar{c}s(d)$ decays of members of the beauty baryon antitriplet to a member of the light baryon octet and a singlet. Under several reasonable assumptions we found $\left\vert \mathcal{A}(\Xi_b^0\rightarrow \Lambda S)/ \mathcal{A}(\Xi_b^0\rightarrow \Xi^0 S)\right\vert\approx 1/\sqrt{6}\, \left\vert V_{cb}^* V_{cd} / (V_{cb}^* V_{cs})\right\vert$ and $\left\vert\mathcal{A}(\Lambda_b\rightarrow \Sigma^0 S)/\mathcal{A}(\Lambda_b\rightarrow \Lambda S)\right\vert \sim 0.02$. These two relations have been recently probed by LHCb for the case of $S=J/\psi$. The former agrees with the measurement, while for the latter our prediction lies close to the upper bound set by LHCb.
\end{abstract} 

\maketitle

\section{Introduction \label{sec:introduction} }

A tremendous amount of $b$-baryons is produced at the LHC~\cite{Cerri:2018ypt}. This 
allows for angular analyses of $\Lambda_b$ decays at LHCb~\cite{Aaij:2013oxa} and ATLAS~\cite{Aad:2014iba} and has led 
to evidence of CP violation in $\Lambda_b$ decays \cite{Aaij:2016cla}. It is now feasible to scrutinize  
rare or suppressed $b$-baryon decays: Recent results include the first observation of 
$\Lambda_b\rightarrow \Lambda\gamma$~\cite{Aaij:2019hhx} and the analysis of the isospin suppressed  
$\Lambda_b\rightarrow \Sigma^0 J/\psi$ decay and the Cabibbo-suppressed 
decay $\Xi_b^0\rightarrow \Lambda J/\psi$~\cite{Aaij:2019hyu}.

These increasingly precise measurements of baryon decays motivate us 
to perform an SU(3)$_F$ analysis of 
$b\rightarrow c\bar{c}q$ (with $q=s,d$) decays
of the heavy $b$-baryon $\mathbf{\overline{3}}$ to the light baryon $\mathbf{8}$ and an SU(3)$_F$ singlet,
$\overline{\mathbf{3}}_b\rightarrow \mathbf{8}_b \otimes \mathbf{1}$.
From the perspective of SU(3)$_F$ it makes no difference if the singlet, which we denote as $S$, is a $J/\psi$ or any final state particle that does not carry any SU(3)$_F$ flavor, for example, a photon or a lepton pair.

We start our analysis using two separate assumptions: 
(1) We work in the SU(3)$_F$ limit and (2) we treat the $\Lambda$ and $\Sigma^0$ as isospin eigenstates.  
We emphasize that these assumptions are not connected to each other. 
We later relax these assumptions and take into account corrections to the SU(3)$_F$ limit as well as deviations of the mass eigenstates of $\Lambda$ and 
$\Sigma^0$ from their isospin eigenstates. 

At leading order the decays $\overline{\mathbf{3}}_b\rightarrow \mathbf{8}_b \otimes \mathbf{1}$ 
are mediated by tree-level $b\rightarrow c\bar{c}q$ transitions. These correspond to a $\mathbf{3}$ operator.
In full generality however, we have to 
take into account additional contributions from loops that generate effective $b\rightarrow t\bar{t} q$ and 
$b\rightarrow u\bar{u} q$ transitions. 
The contribution from  $b\rightarrow t\bar{t} q$ can be neglected as
it is a penguin and therefore suppressed and it gives only another $\mathbf{3}$ under SU(3)$_F$. 
In contrast, the up quarks in $b\rightarrow u\bar{u} q$ can induce intermediate on-shell states leading to nontrivial effects from rescattering.
Specifically, the $b\rightarrow u\bar{u} q$ transition has a more complicated isospin and SU(3)$_F$ 
structure and induces the higher SU(3)$_F$ representations $\mathbf{\overline{6}}$ and $\mathbf{15}$. 
Therefore, as higher SU(3)$_F$ representations stem from rescattering, in the literature it is often assumed that these are suppressed.

Our strategy is to start with a very general model-independent 
viewpoint and then introduce additional assumptions step by step.
While we mainly concentrate in this paper on the case where $S=J/\psi$,
the general nature of our results make it possible to apply them also to radiative and semileptonic decays.

CKM-leading SU(3)$_F$ limit Clebsch-Gordan coefficients for 
$\overline{\mathbf{3}}_b\rightarrow \mathbf{8}_b\otimes \mathbf{1}$ 
in $b\rightarrow s$ transitions have been presented in 
Refs.~\cite{Voloshin:2015xxa, Fayyazuddin:2017sxq, Gutsche:2018utw}.
In Refs.~\cite{Fayyazuddin:2017sxq, Hsiao:2015cda, Hsiao:2015txa, Zhu:2018jet} hadronic models based on QCD factorization have been utilized, 
and in Refs.~\cite{Gutsche:2017wag,Gutsche:2018utw} a covariant confined quark model has been applied.
An SU(3)$_F$ analysis of $b$-baryon antitriplet decays to the light baryon octet and the $\eta_1$ singlet can be found in Ref.~\cite{Roy:2019cky}.

Further applications of SU(3)$_F$ to $b$ baryon decays can be found in Refs.~\cite{Gronau:2013mza, He:2015fsa, He:2015fwa, Arora:1992yq, Du:1994qt, Korner:1994nh}. Works on $b$ baryon decays beyond their SU(3)$_F$ treatment are given in 
Refs.~\cite{Hsiao:2014mua, Gronau:2015jgh, Gronau:2016xiq, Egolf:2002nk, Leibovich:2003tw}.
Applications of SU(3)$_F$ methods on non-$b$ baryon decays can be found in Refs.~\cite{Grossman:2018ptn, Lu:2016ogy, Geng:2018upx, Jia:2019zxi, Savage:1989qr, Savage:1990pr, Wang:2019alu, Geng:2018bow, Geng:2018plk, Geng:2017mxn, Zhao:2018mov, Gronau:2018vei}.
Further literature on baryon decays is given in Refs.~\cite{Ebert:1983ih, Bigi:1981yz, Bigi:2012ev}.
Discussions of baryonic form factors can be found in Refs.~\cite{Isgur:1990pm, Khodjamirian:2011jp, Wang:2015ndk, Husek:2019wmt, Granados:2017cib, Kubis:2000aa, Mannel:2011xg, Detmold:2012vy, Detmold:2016pkz, Bernlochner:2018bfn}.

We present our SU(3)$_F$ analysis 
including isospin and SU(3)$_F$ breaking in Sec.~\ref{sec:group-theory}.
After that we estimate in Sec.~\ref{sec:mixing} the effect of $\Sigma^0$--$\Lambda$ mixing in $\Lambda_b$ decays, which is in general 
scale- and process-dependent, i.e.~non-universal. We compare with recent experimental results in Sec.~\ref{sec:data} 
and conclude in Sec.~\ref{sec:conclusions}.

\section{SU(3)$_F$ Analysis \label{sec:group-theory}}

\subsection{General SU(3)$_F$ Decomposition}

The $b\rightarrow c\bar{c}q$ (with $q=s,d$) decays of $\Lambda_b$, $\Xi_b^-$ and $\Xi_b^0$, which form the heavy baryon $\mathbf{\bar{3}}$, into a singlet $S$ (e.g. $S=J/\psi$, $\gamma$, $l^+l^-$, \dots) and a member of the light baryon $\mathbf{8}$, share a common set of reduced SU(3)$_F$ matrix elements after the application of the  
Wigner-Eckart theorem. These decays are specifically:
\begin{itemize}
\item $b\rightarrow s c\bar{c}$ transitions:
\begin{align}
\Lambda_b \rightarrow  \Lambda  S\,,  & &
\Lambda_b \rightarrow  \Sigma^0 S\,,  & &
\Xi_b^0   \rightarrow  \Xi^0    S\,,  & &
\Xi_b^-   \rightarrow  \Xi^-    S\,. \label{eq:bs-transition} 
\end{align}
\item $b\rightarrow d c\bar{c}$ transitions:
\begin{align}
\Xi_b^0 \rightarrow \Lambda S\,, & &
\Xi_b^0 \rightarrow \Sigma^0 S\,, & &
\Lambda_b \rightarrow n S\,, & & 
\Xi_b^- \rightarrow \Sigma^- S \,. \label{eq:bd-transition} 
\end{align}
\end{itemize}
Note that there are two additional allowed decays $\Lambda_b\rightarrow \Xi^0 J/\psi$ and $\Xi_b^0\rightarrow n J/\psi$ which are however highly suppressed by two insertions of weak effective operators, so we do not consider them in our study here. 
The SU(3)$_F$ quantum numbers and masses are given in Table~\ref{tab:particles}. 
In this section we discuss the SU(3)$_F$ limit, SU(3)$_F$-breaking effects are treated in Sec.~\ref{sec:su3breaking}. 

We can write the SU(3)$_F$ structure of the relevant $b\rightarrow s$ and $b\rightarrow d$ Hamiltonians as~\cite{Zeppenfeld:1980ex}
\begin{align}
\mathcal{H}^{b\rightarrow s}  &=  \lambda_{cs} (\bar{c} b) (\bar{s} c)
				+ \lambda_{us} (\bar{u} b) (\bar{s} u)
				+ \lambda_{ts} (\bar{t} b) (\bar{s} t) \nn\\
				&= \lambda_{cs} \left(\mathbf{3}\right)^c_{0,0,-\frac{2}{3}} +
				 \lambda_{us} \left(
		\left(\mathbf{3}\right)^u_{0,0,-\frac{2}{3}} + \left(\mathbf{\bar{6}}\right)^u_{1,0,-\frac{2}{3}} + \sqrt{6} \left(\mathbf{15}\right)^u_{1,0,-\frac{2}{3}} + \sqrt{3} \left(\mathbf{15}\right)^u_{0,0,-\frac{2}{3}} \right)\,, 
		\label{eq:operator-quark-content} \\
\mathcal{H}^{b\rightarrow d} &= \lambda_{cd} (\bar{c} b) (\bar{d} c) + 
			        \lambda_{ud} (\bar{u} b) (\bar{d} u) + 
			        \lambda_{td} (\bar{t} b) (\bar{d} t) \nn \\ 
				&= \lambda_{cd}  \left(\mathbf{3}\right)^c_{\frac{1}{2},-\frac{1}{2},\frac{1}{3}}  +
				\lambda_{ud} \left( 
			\left(\mathbf{3}\right)^u_{\frac{1}{2},-\frac{1}{2},\frac{1}{3}} -
			\left(\overline{\mathbf{6}}\right)^u_{\frac{1}{2},-\frac{1}{2},\frac{1}{3}} +
			\sqrt{8} \left(\mathbf{15}\right)^u_{\frac{3}{2},-\frac{1}{2},\frac{1}{3}} +
			\left(\mathbf{15}\right)^u_{\frac{1}{2},-\frac{1}{2}, \frac{1}{3}}
			\right)\,. \label{eq:bd-operator-quark-content} 
\end{align}
See also Refs.~\cite{Jung:2012mp} and \cite{Jung:2014jfa} for the application of
these Hamiltonians to $B\rightarrow J/\psi K$ and $B\rightarrow DD$, respectively. 
The notation for the subindices are such that
$\left(\mathbf{N}\right)_{I,I_3,Y}$
refers to the irreducible representation~$\mathbf{N}$ of SU(3)$_F$ using the quantum numbers of strong isospin~$I$, $I_3$
and strong hypercharge~$Y$. In the standard basis of the
Gell-Mann matrices $I_3$ and $Y$ correspond to the eigenvalues of
$\lambda_3$ and $\lambda_8$, respectively. 
We further use the notation 
\begin{align}
\lambda_{cs} &\equiv V_{cb}^* V_{cs}\sim \lambda^2\,, &
\lambda_{us} &\equiv V_{ub}^* V_{us}\sim \lambda^4\,, &
\lambda_{ts} &\equiv V_{tb}^* V_{ts}\sim \lambda^2\,, \label{eq:CKM-hierarchy-1}\\
\lambda_{cd} &\equiv V_{cb}^* V_{cd}\sim \lambda^3\,, &
\lambda_{ud} &\equiv V_{ub}^* V_{ud}\sim \lambda^3\,, &
\lambda_{td} &\equiv V_{tb}^* V_{td}\sim \lambda^3\,, \label{eq:CKM-hierarchy-2}
\end{align}
for the CKM matrix element combinations, where we indicate the hierarchies using the Wolfenstein parameter $\lambda$.

Note that in Eqs.~(\ref{eq:operator-quark-content}) and (\ref{eq:bd-operator-quark-content}) it is understood that SU(3)$_F$ operators in front of different CKM matrix elements have to be differentiated as they stem from different underlying operators. 
For instance, even if the two triplets generate linearly dependent Clebsch-Gordan coefficients, the respective matrix elements themselves are independent. They can, for example, have a relative strong phase.

We write the reduced SU(3)$_F$ limit matrix elements as $A_q^k$\,, where $k$ is the respective SU(3)$_F$ representation 
in the Hamiltonian and $q$ denotes the operator it stems from. 
The initial state is always a $\ket{\overline{\mathbf{3}}}$ and the final state is always a $\ket{\mathbf{8}}$,
so that we are left with four reduced matrix elements in the SU(3)$_F$ limit:
\begin{align}
A^3_c\,, \qquad  
A^3_u\,, \qquad 
A^{\overline{6}}_u\,, \qquad 
A^{15}_u\,. 
\end{align}
The SU(3)$_F$ limit decomposition is given in Table~\ref{tab:SU3}.
The CKM-leading part of the $b\rightarrow s$ transitions agrees with Refs.~\cite{Voloshin:2015xxa,Fayyazuddin:2017sxq, Gutsche:2018utw}.
The Clebsch-Gordan coefficients are obtained using Refs.~\cite{deSwart:1963pdg,Kaeding:1995vq,Kaeding:1995re}.
The normalization of the amplitudes is such that 
\begin{align}
\mathcal{B}(B_1\rightarrow B_2 S) &= \vert \mathcal{A}(B_1\rightarrow B_2 S)\vert^2 \times \mathcal{P}(B_1, B_2, S) \,, 
\label{eq:phasespace}
\end{align}
with the two-body decay phase space factors 
\begin{align}
\mathcal{P}(B_1, B_2, S) &\equiv \frac{ \tau_{B_1}}{16 \pi m_{B_1}^3} 
	\sqrt{  (m_{B_1}^2 - (m_{B_2} - m_S)^2) ( m_{B_1}^2 - (m_{B_2} + m_S)^2) }\,.
\label{eq:PSfactor}
\end{align}
Note that in cases where the SU(3)$_F$ singlet $S$ is a multibody state, e.g.~$S=l^+l^-$,
we imply the appropriate phase space integration in Eq.~(\ref{eq:phasespace}).
Note further, that we still work in the SU(3)$_F$ limit of the decay
amplitudes. Eq.~(\ref{eq:PSfactor}) only accounts for the trivial SU(3)$_F$ breaking from phase space effects. Additional SU(3)$_F$ breaking contributions are
discussed in Sec.~\ref{sec:su3breaking}.
Therein, we estimate SU(3)$_F$ breaking effects to be of order 20\%.
Note that the amplitudes in Eq.~(\ref{eq:phasespace}) have a mass dimension, but we
always care about ratios, so we can think about them as dimensionless quantities.
Note that phase space effects are of order $3\%$ and thus they are
well within the errors and could or could not be taken into account.
For a
model-dependent way to estimate these effects one can, for example, employ
form factor results in Refs.~\cite{Gutsche:2013oea, Gutsche:2018utw}.

\begin{figure}[t]
 \begin{center}
  \subfigure[\, $C$]{\includegraphics[width=0.4\textwidth]{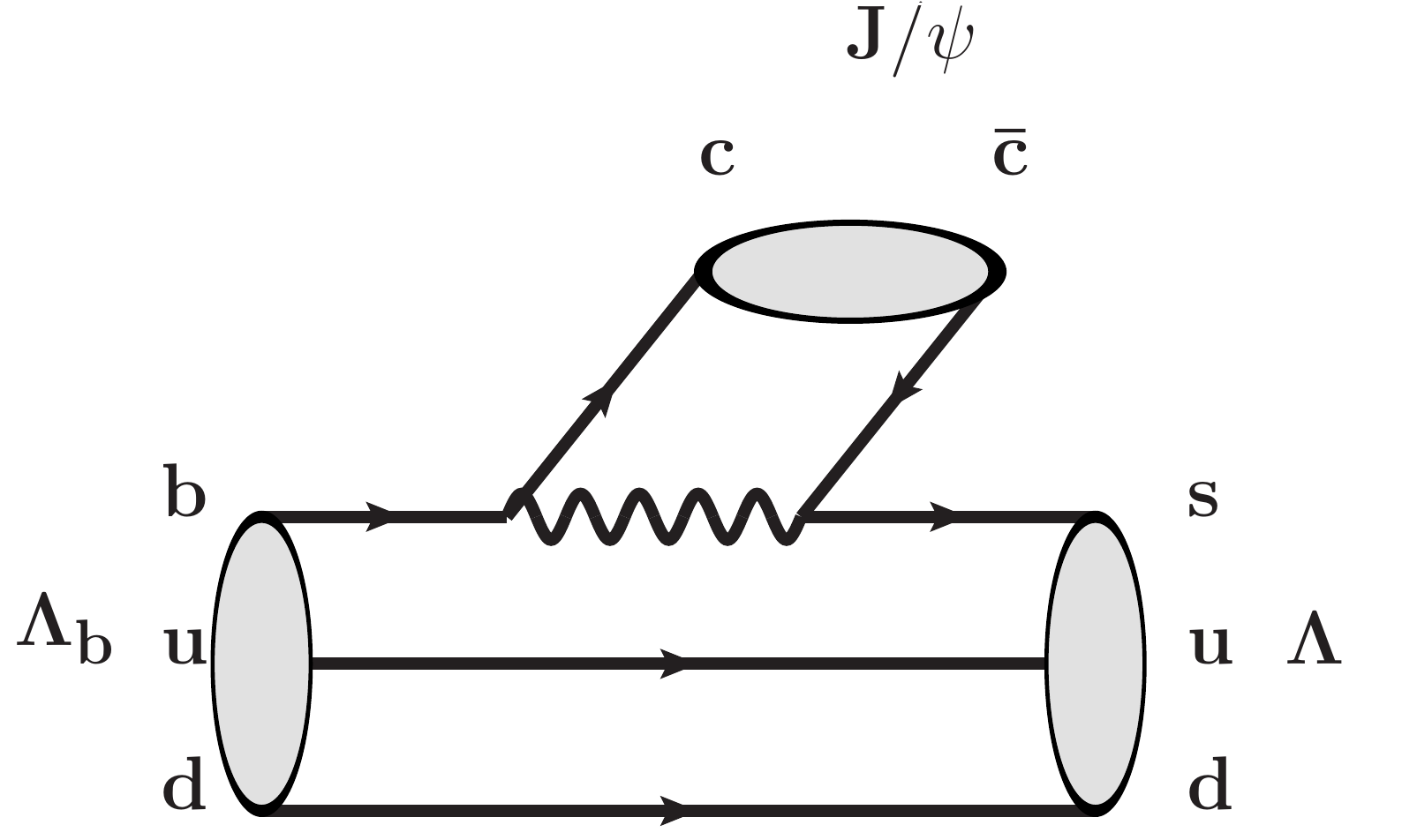}}
  \subfigure[\, $E$]{\includegraphics[width=0.4\textwidth]{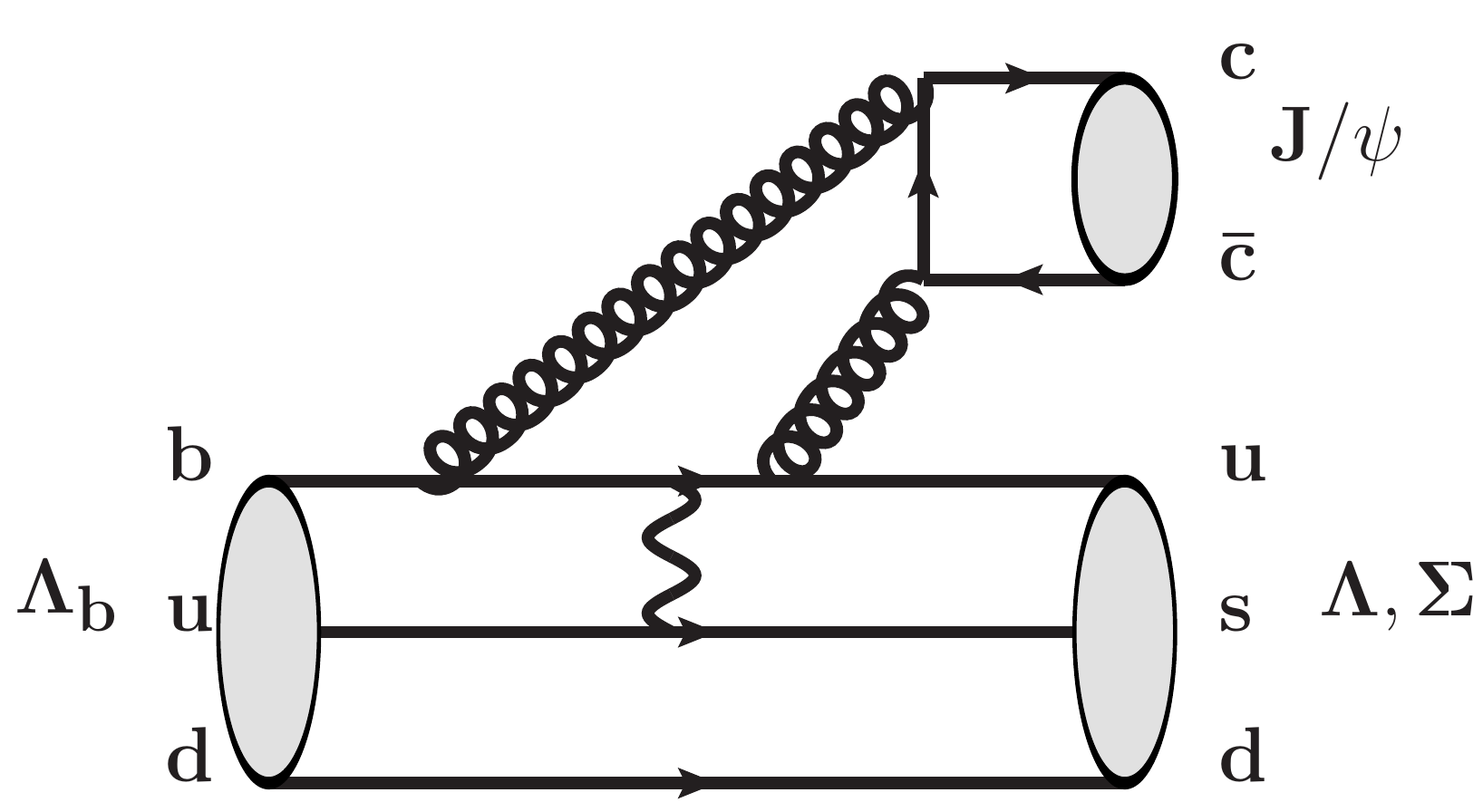}}
  \subfigure[\, $P_q$]{\includegraphics[width=0.4\textwidth]{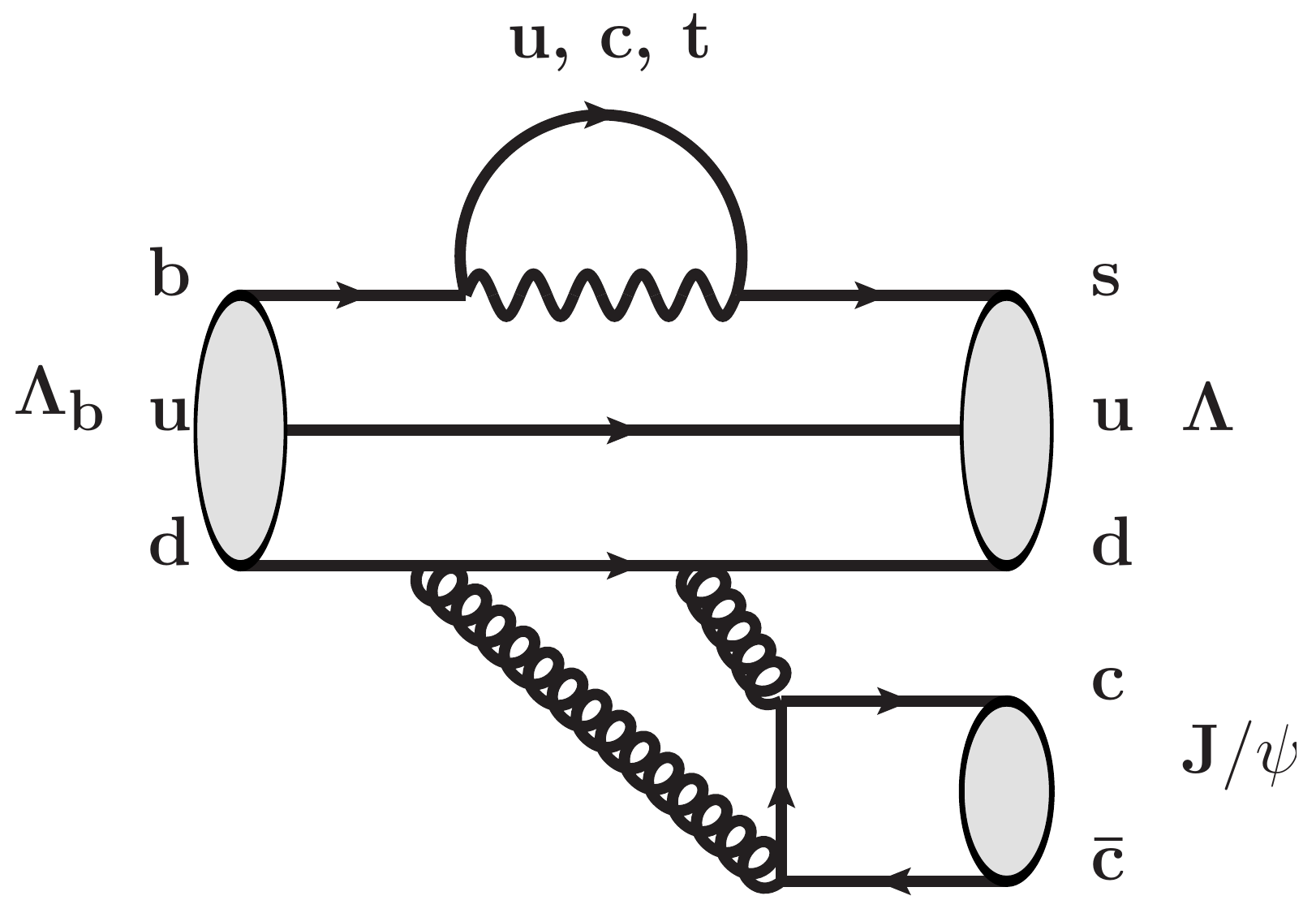}}
  \caption{Topological diagrams for the decays $\Lambda_b\rightarrow \Lambda J/\psi$ and $\Lambda_b\rightarrow \Sigma J/\psi$. Note that in the exchange diagram one gluon alone can not create the $J/\psi$ because it is a color singlet. 
\label{fig:topo}}
\end{center}
\end{figure}

The reduced SU(3)$_F$ matrix elements can in principle be matched on 
a color suppressed tree diagram $C$, an exchange diagram $E$ and penguin diagrams $P_q$ with quark $q$ running in the loop. 
As examples we show the topological diagrams for $\Lambda_b\rightarrow \Lambda J/\psi$ and 
$\Lambda_b\rightarrow \Sigma^0 J/\psi $ in Fig.~\ref{fig:topo}. In the following, however, we only perform
the group theory treatment.

The combined matrix of Clebsch-Gordan coefficients of $b\rightarrow s$ and $b\rightarrow d$ decays in Table~\ref{tab:SU3} has matrix rank four, i.e., there are four sum rules, which read  
\begin{align}
-\sqrt{\frac{3}{2}}\mathcal{A}(\Lambda_b\rightarrow \Lambda S) +
\frac{1}{\sqrt{2}} \mathcal{A}(\Lambda_b\rightarrow \Sigma^0 S ) +
 \mathcal{A}(\Xi_b^0\rightarrow \Xi^0 S ) &= 0\,,& \text{(SU(3)$_F$ sum rule)}  \\
\sqrt{\frac{3}{2}}\mathcal{A}(\Xi_b^0\rightarrow \Lambda S) 
-\frac{1}{\sqrt{2}}\mathcal{A}(\Xi_b^0\rightarrow \Sigma^0 S) +
\mathcal{A}(\Lambda_b\rightarrow n S) &= 0\,,& \text{(SU(3)$_F$ sum rule)} \\ 
-\sqrt{2}  \mathcal{A}(\Lambda_b\rightarrow \Sigma^0 S) \frac{\lambda_{ud}}{\lambda_{us}}
+\sqrt{6} \mathcal{A}(\Xi_b^0 \rightarrow \Lambda S)
+ \mathcal{A}(\Lambda_b\rightarrow nS) &= 0\,, & \text{(SU(3)$_F$ sum rule)} \\
\sqrt{\frac{3}{2}}  \mathcal{A}(\Lambda_b\rightarrow \Lambda S)  \frac{\lambda_{ud}}{\lambda_{us}}
-\frac{3}{\sqrt{2}}  \mathcal{A}(\Lambda_b\rightarrow \Sigma^0 S) \frac{\lambda_{ud}}{\lambda_{us}} &\nn\\ 
- \mathcal{A}(\Xi_b^- \rightarrow \Xi^- S) \frac{\lambda_{ud}}{\lambda_{us}}
+\sqrt{6} \mathcal{A}(\Xi_b^0\rightarrow \Lambda S) 
+\mathcal{A}(\Xi_b^-\rightarrow \Sigma^- S)
&= 0\,, & \text{(SU(3)$_F$ sum rule)} 
\end{align}
all of which are SU(3)$_F$ sum rules, and there is no isospin sum rule.
Note that there are two sum rules which mix $b\rightarrow s$ and $b\rightarrow d$ decays and two which do not.
These sum rules are valid in the SU(3)$_F$ limit irrespective of the power counting of the CKM matrix elements, assumptions on the reduced matrix elements,
or the particular SU(3)$_F$ singlet $S$, i.e. they are completely generic.  

\subsection{Assumptions on CKM Hierarchy and Rescattering}

We now make some assumptions, which are not completely generic, i.e. their validity can for example depend on the particular considered SU(3)$_F$ singlet $S$, e.g. if $S=J/\psi$ or $S=\gamma$.

We first neglect the CKM-suppressed
amplitude in $b\rightarrow s$ decays, that is we set 
$\lambda_{us}/\lambda_{cs}\to 0$. In
the isospin and SU(3)$_F$ limit for $b\rightarrow s$ decays we have then only one contributing reduced matrix element:
\begin{align}
\mathcal{A}(\Lambda_b\rightarrow \Sigma^0 S) &= 0\,, & \text{(isospin sum rule)} \label{eq:CKMleading-isospin-sigma} \\
	\mathcal{A}(\Xi_b^0\rightarrow \Xi^0 S) &= \mathcal{A}(\Xi_b^-\rightarrow \Xi^- S)\,, & \text{(isospin sum rule)} \label{eq:sumrules-1} \\  
	\mathcal{A}(\Xi_b^0\rightarrow \Xi^0 S) &= \sqrt{\frac{3}{2}}\mathcal{A}(\Lambda_b\rightarrow \Lambda S)\,.  & \text{(SU(3)$_F$ sum rule)} \label{eq:sumrules-1a}
\end{align}

We now move to make another assumption and that is to also neglect the
$\lambda_{ud}$ terms for the $b\rightarrow d$ transitions. Despite the formal power counting Eq.~(\ref{eq:CKM-hierarchy-2}), 
that is $\vert \lambda_{ud}\vert \simeq \vert \lambda_{cd}\vert$, numerically we actually have~\cite{Tanabashi:2018oca}
\begin{align}
\left\vert \frac{\lambda_{ud}}{\lambda_{cd}}\right\vert &\approx 0.38\,.
\end{align} 
Moreover, it is plausible that $A_u^{\overline{6}}$ and $A_u^{15}$ are suppressed because they result from light quarks stemming from $b\rightarrow u\bar{u}s(d)$ which induce intermediate on-shell states that rescatter into $c\bar c$, see also Refs.~\cite{Hernandez:1994rh, Gronau:1994bn, Gronau:1994rj, Gronau:1995hm, Neubert:1998jq}. 
Under the assumption that these terms are more or equally suppressed as SU(3)$_F$-breaking
effects we have many more relations. All seven non-zero 
decays we considered in Table~\ref{tab:SU3} are then simply related by the Clebsch-Gordan coefficients in the first
column. In addition to the sum rules Eqs.~(\ref{eq:CKMleading-isospin-sigma})--(\ref{eq:sumrules-1a}), we have then
\begin{align}
\sqrt{2}\, \mathcal{A}(\Xi_b^0\rightarrow \Sigma^0 S )   &= \mathcal{A}(\Xi_b^-\rightarrow \Sigma^- S) \,, & \text{(isospin sum rule)} \\
\mathcal{A}(\Xi_b^0\rightarrow \Xi^0 S) &= -\sqrt{6}\, \mathcal{A}(\Xi_b^0\rightarrow \Lambda S) \frac{\lambda_{cs}}{\lambda_{cd}} \,,& \text{(SU(3)$_F$ sum rule)} \label{eq:key-sum-rule}\\
-\sqrt{6}\, \mathcal{A}(\Xi_b^0\rightarrow \Lambda S)  &= \sqrt{2}\, \mathcal{A}(\Xi_b^0\rightarrow \Sigma^0 S )   \,,& \text{(SU(3)$_F$ sum rule)} \\ 
\sqrt{2}\, \mathcal{A}(\Xi_b^0\rightarrow \Sigma^0 S )    &= \mathcal{A}(\Lambda_b\rightarrow n S)   \,,& \text{(SU(3)$_F$ sum rule)} \\
\mathcal{A}(\Lambda_b\rightarrow n S)   &=  \mathcal{A}(\Xi_b^-\rightarrow \Sigma^- S) \,.& \text{(SU(3)$_F$ sum rule)} 
\end{align}

\subsection{Isospin and $U$-Spin Decompositions}

For comprehensiveness, we give also the isospin and $U$-spin decompositions of the Hamiltonians, which read 
\begin{align}
\mathcal{H}_{b\rightarrow s} &= \lambda_{cs} (0,0)^c_{I} +
           \lambda_{us} \left( (0,0)^u_{I} + (1,0)^u_{I} \right) \label{eq:H-iso-bs}\\ 
        &= \lambda_{cs} \left(\frac{1}{2},-\frac{1}{2}\right)^c_{U} +
           \lambda_{us} \left(\frac{1}{2},-\frac{1}{2}\right)^u_{U} \,, \label{eq:H-u-bs}
\end{align}
and
\begin{align}
\mathcal{H}_{b\rightarrow d} &= \lambda_{cd} \left(\frac{1}{2},-\frac{1}{2}\right)^c_{I} +
           \lambda_{ud} \left( \left(\frac{3}{2},-\frac{1}{2}\right)^u_{I} + \left(\frac{1}{2},-\frac{1}{2}\right)^u_{I} \right) \label{eq:H-iso-bd}\\
        &= \lambda_{cd}  \left(\frac{1}{2},\frac{1}{2}\right)^c_{U} +
           \lambda_{ud} \left(\frac{1}{2},\frac{1}{2}\right)^u_{U}\,, \label{eq:H-u-bd} 
\end{align}
where we use the notation 
\begin{align}
\left(i,j\right)_I^q \equiv \mathcal{O}^{\Delta I=i}_{\Delta I_3=j}\,, \qquad
\left(i,j\right)_U^q \equiv \mathcal{O}^{\Delta U=i}_{\Delta U_3=j}\,,
\end{align}
where $q$ denotes the quark content of the operator the representation stems from and we absorbed Clebsch-Gordan coefficients into operators.

Using the isospin and $U$-spin states in Table~\ref{tab:particles}, we obtain the isospin decompositions in Tables~\ref{tab:isospin1} and \ref{tab:isospin2} and the $U$-spin decomposition in Table~\ref{tab:uspin}.
We note that the SU(3)$_F$ decomposition includes more information than the isospin and $U$-spin tables each on their own.
An example is the ratio
\begin{align}
\left\vert \frac{\mathcal{A}(\Xi_b^0\rightarrow \Lambda S)}{\mathcal{A}(\Xi_b^0\rightarrow \Xi^0 S)}\right\vert = 
 \frac{1}{\sqrt{2}}  \left\vert  \frac{\bra{0}\frac{1}{2}\ket{\frac{1}{2}} }{\bra{\frac{1}{2}}0\ket{\frac{1}{2}}}\right\vert \left\vert\frac{\lambda_{cd}}{\lambda_{cs} }\right\vert  \,.
\end{align}
where the appearing reduced matrix elements are not related, e.g. the final states belong to different isospin representations.
That means we really need SU(3)$_F$ to find the relation Eq.~(\ref{eq:key-sum-rule}). 

We can make this completely transparent by writing out the implications of Eq.~(\ref{eq:CKMleading-isospin-sigma}) 
for the corresponding $U$-spin decomposition.
From Table~\ref{tab:uspin} and Eq.~(\ref{eq:CKMleading-isospin-sigma}) it follows for the $U$-spin matrix elements
\begin{align}
-\frac{\sqrt{3}}{2 \sqrt{2}} \left\langle 0 \left\vert\frac{1}{2}\right\vert \frac{1}{2}\right\rangle^c + 
	\frac{1}{2\sqrt{2}} \left\langle 1\left\vert \frac{1}{2}\right\vert \frac{1}{2}\right\rangle^c &= 0\,.
\end{align}
Inserting this relation into the $U$-spin decomposition of the decay $\Xi_b^0\rightarrow \Lambda S$ in Table~\ref{tab:uspin}, we obtain
\begin{align}
\mathcal{A}(\Xi_b^0\rightarrow \Lambda S) &=  \frac{1}{\sqrt{6}}  \lambda_{cd}
	\left\langle 1\left\vert \frac{1}{2}\right\vert \frac{1}{2}\right\rangle^c \,. \label{eq:Xib-LambdaS-U-Spin}
\end{align}
Comparing this expression with the $U$-spin decomposition of the decay $\Xi_b^0\rightarrow \Xi^0 S$ in Table~\ref{tab:uspin}, we arrive again at the sum rule 
Eq.~(\ref{eq:key-sum-rule}). 

In order that Eq.~(\ref{eq:key-sum-rule}) holds we need not only the suppression of other 
SU(3)$_F$ limit contributions as discussed above, but also the suppression of both isospin and $U$-spin violating 
contributions. A non-vanishing dynamic isospin breaking contribution to $\Lambda_b\rightarrow \Sigma^0 S$ 
would also be reflected in isospin and SU(3)$_F$-breaking violations of Eq.~(\ref{eq:key-sum-rule}). 
We make this correlation explicit in Sec.~\ref{sec:su3breaking}.

\subsection{CP Asymmetry Sum Rules}

Due to a general sum rule theorem given in Ref.~\cite{Gronau:2000zy} that relates direct CP asymmetries of decays connected by a complete interchange of $d$ and $s$ quarks~\cite{Gronau:2000zy,Fleischer:1999pa, Gronau:2000md, He:1998rq}, we can directly write down two $U$-spin limit sum rules: 
\begin{align}
\frac{
a_{CP}^{\mathrm{dir}}(\Xi_b^0 \rightarrow \Xi^0 S)
}{
a_{CP}^{\mathrm{dir}}(\Lambda_b \rightarrow n S)
} &= - \frac{\tau(\Xi_b^0)}{\tau(\Lambda_b) }\frac{
\mathcal{B}(\Lambda_b \rightarrow n S)
}{
\mathcal{B}(\Xi_b^0 \rightarrow \Xi^0 S)
}\,, \label{eq:CP-uspin-1} \\
\frac{
a_{CP}^{\mathrm{dir}}(\Xi_b^-   \rightarrow  \Xi^- S)
}{
a_{CP}^{\mathrm{dir}}(\Xi_b^- \rightarrow \Sigma^- S)
} &= -\frac{
\mathcal{B}( \Xi_b^- \rightarrow \Sigma^- S)
}{
\mathcal{B}( \Xi_b^- \rightarrow  \Xi^- S)
}\,, \label{eq:CP-uspin-2}
\end{align}
where the branching ratios imply CP averaging. 
Note that the general U-spin rule leading to Eqs.~(\ref{eq:CP-uspin-1}) and (\ref{eq:CP-uspin-2}) 
also applies to multi-body final states, as pointed out in Refs.~\cite{Gronau:2000zy, Bhattacharya:2013boa, Grossman:2018ptn}.
It follows that Eqs.~(\ref{eq:CP-uspin-1}) and (\ref{eq:CP-uspin-2}) apply also when $S$ is a multi-body 
state like $S=l^+l^-$.

Note that although the quark content of the $\Lambda$ and $\Sigma$ is $uds$, this does not mean that a complete interchange of $d$ and $s$ quarks gives the identity. The reason is given by the underlying quark wave functions~\cite{Isgur:1979ed} 
\begin{align}
\ket{\Lambda} &\sim \frac{1}{\sqrt{2}} \left(ud - du\right)s\,, \qquad
\ket{\Sigma^0}   \sim  \frac{1}{\sqrt{2}} \left(ud + du\right)s\,, \label{eq:lambda-sigma-quark-wv}
\end{align}
where we do not write the spin wave function. Eq.~(\ref{eq:lambda-sigma-quark-wv}) shows explicitly that a complete interchange of $d$ and $s$ quarks in $\Lambda$ or $\Sigma^0$ does not result again in a $\Lambda$ or $\Sigma^0$ wave function, respectively. 
This is similar to the situation for $\eta$ and $\eta'$, where no respective particles correspond to a complete interchange of $d$ and $s$ quarks~\cite{Wang:2019dls}, see e.g.~the quark wave functions given in Ref.~\cite{Bhattacharya:2009ps}. 

We can put this into a different language, namely that in the $U$-spin basis the large mixing of $\ket{1,0}_U$ and $\ket{0,0}_U$ to the $U$-spin states of $\Lambda$ and $\Sigma^0$, see Table \ref{tab:particles}, destroys two sum rules which exist for the $U$-spin eigenstates. To be explicit, 
we define $U$-spin eigenstates which are not close to mass eigenstates
\begin{align}
\ket{X} &= \ket{0,0}_U\,, \quad \ket{Y} = \ket{1,0}_U\,. \label{eq:hypothetical}
\end{align}
For these, we obtain the $U$-spin decomposition given in Table~\ref{tab:uspin-hypo}. 
From that it is straightforward to obtain another two CP asymmetry sum rules. These are however impractical, 
because there is no method available to prepare $\Lambda$ and $\Sigma^0$ as $U$-spin eigenstates, instead of 
approximate isospin eigenstates.
Consequently, we are left only with the two CP asymmetry sum rules Eqs.~(\ref{eq:CP-uspin-1}) and (\ref{eq:CP-uspin-2}).

Note that CKM-leading SU(3)$_F$ breaking by itself cannot contribute to CP violation, because it comes only with relative strong phases but not with the necessary relative weak phase. 
Therefore, the individual CP asymmetries can be written as 
\begin{align}
a_{CP}^{\mathrm{dir}} &= \mathrm{Im}\frac{\lambda_{uq}}{\lambda_{cq}} \mathrm{Im}\frac{A_u}{A_c}\,, 
\end{align}
where $A_{u,c}$ have only a strong phase and to leading order in Wolfenstein-$\lambda$ we have~\cite{Tanabashi:2018oca}
\begin{align}
\mathrm{Im}\left(\frac{\lambda_{us}}{\lambda_{cs}}\right) &\approx \lambda^2 \bar{\eta} \approx 0.02   \,,
\qquad  
\mathrm{Im}\left(\frac{\lambda_{ud}}{\lambda_{cd}}\right) \approx \bar{\eta} \approx 0.36  \,. \label{eq:aCP-CKM}
\end{align}
Additional suppression from rescattering implies that on top of Eq.~(\ref{eq:aCP-CKM}) we have $\vert A_u\vert \ll \vert A_c\vert$, i.e.~the respective imaginary part is also expected to be small.
This implies that we do not expect to see a nonvanishing CP asymmetry in these decays any time soon. 
The other way around, this prediction is also a test of our assumption that the $\lambda_{uq}$-amplitude is suppressed.

\subsection{SU(3)$_F$ Breaking \label{sec:su3breaking} } 

We consider now isospin and SU(3)$_F$ breaking effects in the CKM-leading part of the $b\rightarrow s$ and $b\rightarrow d$ 
Hamiltonians.
This will become useful once we have measurements of several $b$-baryon decays that are precise enough to see deviations from the 
SU(3)$_F$ limit sum rules.
SU(3)$_F$ breaking effects for charm and beauty decays have been discussed in the literature for a long time~\cite{Hiller:2012xm, Grossman:2019xcj, Grossman:2018ptn, Muller:2015rna, Muller:2015lua, Brod:2012ud, Grossman:2012eb, Grossman:2006jg, Falk:2001hx, Jung:2012mp, Savage:1991wu, Hinchliffe:1995hz, Jung:2009pb, Gronau:2006eb, Pirtskhalava:2011va, Buccella:2019kpn, Brod:2012ud, Grossman:2012ry, Jung:2009pb, Hiller:2012xm}.
They are generated through the spurion 
\begin{align}
& \vect{
\frac{m_u}{\Lambda} - \frac{2}{3} \alpha & 0 & 0 \\
0 & \frac{m_d}{\Lambda} + \frac{1}{3} \alpha & 0 \\
0 & 0 & \frac{m_s}{\Lambda} + \frac{1}{3} \alpha
} \nn\\
&= \frac{1}{3}\frac{m_u + m_d + m_s}{\Lambda} \mathbbm{1}  - 
   \frac{1}{2}\left(\frac{ m_d - m_u }{\Lambda} + \alpha \right) \lambda_3 + 
   \frac{1}{2\sqrt{3}}\left(\frac{m_u + m_d - 2 m_s }{\Lambda} - \alpha  \right)\lambda_8\,, \label{eq:spurion}
\end{align}
with the unity $\mathbbm{1}$ and the Gell-Mann matrices $\lambda_3$ and $\lambda_8$. 
The part of Eq.~(\ref{eq:spurion}) that is proportional to $\mathbbm{1}$ can be absorbed into the SU(3)$_F$ limit part.
It follows that the isospin and SU(3)$_F$-breaking tensor operator is given as   
\begin{align}
\delta\, (\mathbf{8})_{1,0,0} + \varepsilon \, (\mathbf{8})_{0,0,0}\,,
\end{align}
with 
\begin{align}
\delta &= \frac{1}{2}\left(\frac{ m_d - m_u }{\Lambda} + \alpha \right)\,,   \qquad 
\varepsilon  = \frac{1}{2\sqrt{3}}\left(\frac{m_u + m_d - 2 m_s }{\Lambda}-\alpha\right)  \, , \label{eq:iso-and-su3-breaking}
\end{align}
where $\alpha$ is the electromagnetic coupling and we generically
expect the size of isospin and SU(3)$_F$ breaking to be $\delta\sim 1\%$ and $\varepsilon\sim 20\%$, respectively.
Note that the scale-dependence of the quark masses, as well as the
fact that we do not know how to define the scale $\Lambda$ make it
impossible to quote decisive values for $\delta$ and $\varepsilon$. Eventually,
they will have to be determined experimentally for each process of interest separately as they are not universal.

For the tensor products of the perturbation with the CKM-leading SU(3)$_F$ limit operator it follows:
\begin{align}
 (\mathbf{8})_{1,0,0} \otimes (\mathbf{3})^c_{0,0,-\frac{2}{3}}    &= 
		\sqrt{\frac{1}{2}}\left(\mathbf{\overline{6}}\right)_{1,0,-\frac{2}{3}} + 
		\sqrt{\frac{1}{2}}\left(\mathbf{15}\right)_{1,0,-\frac{2}{3}}\,,  \\
 (\mathbf{8})_{0,0,0} \otimes \left(\mathbf{3}\right)^c_{0,0,-\frac{2}{3}}    &= 
		\frac{1}{2} \left(\mathbf{3}\right) _{0, 0 ,-\frac{2}{3}} +
		\frac{\sqrt{3}}{2} \left(\mathbf{15}\right)_{0, 0 ,-\frac{2}{3}}\,, \\
 (\mathbf{8})_{1,0,0} \otimes (\mathbf{3})^c_{\frac{1}{2},-\frac{1}{2},\frac{1}{3}}    &=
			\frac{\sqrt{3}}{4} \left(\mathbf{3}\right)_{\frac{1}{2},-\frac{1}{2},\frac{1}{3}}
			-\sqrt{\frac{1}{8}} \left(\mathbf{\overline{6}}\right)_{\frac{1}{2},-\frac{1}{2},\frac{1}{3}} \nn\\&\quad
			-\sqrt{\frac{1}{48}} \left(\mathbf{15}\right)_{\frac{1}{2},-\frac{1}{2},\frac{1}{3}} 
			+\sqrt{\frac{2}{3}} \left(\mathbf{15}\right)_{\frac{3}{2},-\frac{1}{2},\frac{1}{3}}\,, \\
 (\mathbf{8})_{0,0,0} \otimes \left(\mathbf{3}\right)^c_{ \frac{1}{2},-\frac{1}{2},\frac{1}{3} }    &=
		-\frac{1}{4} \left(\mathbf{3}\right)_{\frac{1}{2},-\frac{1}{2},\frac{1}{3}} 
		-\sqrt{\frac{3}{8}} \left(\mathbf{\overline{6}}\right)_{\frac{1}{2},-\frac{1}{2},\frac{1}{3}}
	    	+\frac{3}{4} \left(\mathbf{15}\right)_{\frac{1}{2},-\frac{1}{2},\frac{1}{3}}\,,
\end{align}
so that we arrive at the SU(3)$_F$ breaking Hamiltonians 
\begin{align}
\mathcal{H}_X^{b\rightarrow s} &\equiv \lambda_{cs}\, \delta \left( \sqrt{\frac{1}{2}}\left(\mathbf{\overline{6}}\right)_{1,0,-\frac{2}{3}} + 
		\sqrt{\frac{1}{2}}\left(\mathbf{15}\right)_{1,0,-\frac{2}{3}} \right) + \nn\\ 
&\quad		 \lambda_{cs}\, \varepsilon \left( \frac{1}{2} \left(\mathbf{3}\right)_{0, 0 ,-\frac{2}{3}} +
		\frac{\sqrt{3}}{2} \left(\mathbf{15}\right)_{0, 0 ,-\frac{2}{3}} \right)\,, \\
 \mathcal{H}_X^{b\rightarrow d} &\equiv \lambda_{cd}\, \delta \left( \frac{\sqrt{3}}{4} \left(\mathbf{3}\right)_{\frac{1}{2},-\frac{1}{2},\frac{1}{3}}
			-\sqrt{\frac{1}{8}} \left(\mathbf{\overline{6}}\right)_{\frac{1}{2},-\frac{1}{2},\frac{1}{3}}
			-\sqrt{\frac{1}{48}} \left(\mathbf{15}\right)_{\frac{1}{2},-\frac{1}{2},\frac{1}{3}}
			+\sqrt{\frac{2}{3}} \left(\mathbf{15}\right)_{\frac{3}{2},-\frac{1}{2},\frac{1}{3}}\right) \nn\\
&\quad			+ \lambda_{cd}\, \varepsilon \left( -\frac{1}{4} \left(\mathbf{3}\right)_{\frac{1}{2},-\frac{1}{2},\frac{1}{3}} 
		-\sqrt{\frac{3}{8}} \left(\mathbf{\overline{6}}\right)_{\frac{1}{2},-\frac{1}{2},\frac{1}{3}}
	    	+\frac{3}{4} \left(\mathbf{15}\right)_{\frac{1}{2},-\frac{1}{2},\frac{1}{3}}\right)\,.
\end{align}
This gives rise to three additional matrix elements
\begin{align}
B^{3}\,, \qquad
B^{\overline{6}}\,, \qquad
B^{15}\,.
\end{align}
The CKM-leading decomposition for $b\rightarrow s$ and $b\rightarrow d$ decays including isospin and SU(3)$_F$ breaking 
is given in Table~\ref{tab:SU3breaking}. The complete $4\times 4$ matrix of the $b\rightarrow s$ matrix has rank four, i.e.~there is no $b\rightarrow s$ sum rule to this order. 
As discussed in Sec.~\ref{sec:group-theory} after Eq.~(\ref{eq:Xib-LambdaS-U-Spin}) we see from Table~\ref{tab:SU3breaking} explicitly that isospin breaking contributions to $\mathcal{A}(\Lambda_b\rightarrow \Sigma^0 S)$ lead at the same time to a deviation of the ratio 
$\vert \mathcal{A}(\Xi_b^0\rightarrow \Lambda S) \vert / \vert \mathcal{A}(\Xi_b^0\rightarrow \Xi^0 S) \vert$ from the result Eq.~(\ref{eq:key-sum-rule}). \\ 

Comparing to results present in the literature, in Ref.~\cite{Roy:2019cky} two separate coefficient matrices of $b\rightarrow s$ and $b\rightarrow d$ decays are 
given in terms of the isoscalar coefficients, i.e.~where the isospin quantum number is still kept in the corresponding 
reduced matrix element.
We improve on that by giving instead the SU(3)$_F$ Clebsch-Gordan coefficient table that makes transparent the 
corresponding sum rules in a direct way and furthermore reveals directly the correlations between $b\rightarrow s$ and 
$b\rightarrow d$ decays. We also find the complete set of sum rules, and discuss how further assumptions lead to additional
sum rules. 
We note that the first two sum rules in Eq.~(43) in Ref.~\cite{Roy:2019cky} are sum rules for coefficient matrix vectors but do not apply to the corresponding amplitudes because of the different CKM factors involved.

\section{$\Sigma^0$--$\Lambda$ Mixing in $\Lambda_b$ decays \label{sec:mixing}}

\subsection{General Considerations}

In this section we study the ratio 
\begin{align}
R\equiv \frac{\mathcal{A}(\Lambda_b\rightarrow \Sigma^0_{\mathrm{phys}} J/\psi)}{\mathcal{A}(\Lambda_b\rightarrow \Lambda_{\mathrm{phys}} J/\psi)} 
		  &= \frac{
			\bra{J/\psi \Sigma^0_{\mathrm{phys}}} \mathcal{H} \ket{\Lambda_b}
			}{
			\bra{J/\psi \Lambda_{\mathrm{phys}}} \mathcal{H} \ket{\Lambda_b}
			}\,. \label{eq:sigma-lambda-MEs}
\end{align}
In order to do this we need the matrix elements appearing in Eq.~(\ref{eq:sigma-lambda-MEs}). 
In the limit where isospin is a good symmetry and
$\Sigma^0_{\mathrm{phys}}$ is an isospin eigenstate, $R$
vanishes, and therefore we are
interested in the deviations from that limit. We study leading order
effects in isospin breaking.

We first note that we can neglect the deviation of $\Lambda_b$ from
its isospin limit. The reason is that 
regarding the mixing of heavy baryons, for example $\Sigma_b$--$\Lambda_b$, $\Xi_c^0$--$\Xi^{'0}_c$ or $\Xi_c^+$--$\Xi^{'+}_c$,
in the quark model one obtains a suppression of the mixing angle with the heavy quark mass~\cite{Copley:1979wj, Maltman:1980er, Franklin:1981rc, Savage:1989jx,Boyd:1996cd, Karliner:2008sv}. 
It follows that for our purposes we can safely neglect the mixing between $\Lambda_b$ and $\Sigma_b$ as it is not only 
isospin suppressed but on top suppressed by the $b$ quark mass. 

We now move to discuss the mixing of the light baryons. 
It has already been pointed out in Ref.~\cite{Boyd:1996cd}, that a
description with a single mixing angle captures only part of the
effect. The reason is 
because isospin breaking contributions will affect not only the mixing
between the states but also the decay amplitude.
The non-universality is also reflected in the fact that the
$\Lambda_b\rightarrow \Sigma^0$ transition amplitude vanishes in the heavy quark
limit at large recoil, i.e.~in the phase space when $\Sigma^0$ carries
away a large fraction of the energy \cite{Mannel:2011xg}, see also
Ref.~\cite{Leibovich:2003tw} for the heavy quark limit of similar
classes of decays. 

To leading order in isospin breaking we consider two effects, the
mixing between $\Lambda$ and $\Sigma^0$ as well as the correction to the
Hamiltonian. We discuss these two effects below.

Starting with the wave function mixing angle $\theta_m$\,, this is defined as the mixing angle  
between the isospin limit states 
$\ket{\Sigma^0} = \ket{1,0}_I$ and $\ket{\Lambda} = \ket{0,0}_I$, see Eq.~(\ref{eq:lambda-sigma-quark-wv}), into the physical states (see Refs.~\cite{Coleman:1961jn, Dalitz:1964es, Horsley:2014koa, Gal:2015iha, Horsley:2015lha, Kordov:2019oer})
\begin{align}
\ket{\Lambda_{\mathrm{phys}}} &= \cos\theta_m \ket{\Lambda} - \sin\theta_m \ket{\Sigma^0}\,, \label{eq:mix-angle-def-1} \\
\ket{\Sigma^0_{\mathrm{phys}}} &= \sin\theta_m \ket{\Lambda} + \cos\theta_m \ket{\Sigma^0}\,. \label{eq:mix-angle-def-2}
\end{align}
The effect stems from the non-vanishing mass difference $m_d-m_u$ as well as different 
electromagnetic charges \cite{Isgur:1979ed} which lead to a hyperfine mixing between the isospin limit states. 
A similar mixing effect takes place for the light mesons in form of singlet 
octet mixing of $\pi^0$ and $\eta^{(')}$~\cite{Feldmann:2002kz, Feldmann:1998sh, Feldmann:1998vh, Feldmann:1997vc, Dudek:2011tt, Ottnad:2019xmh}. 

As for the Hamiltonian, we write $\mathcal{H}=\mathcal{H}_0+\mathcal{H}_1$ where $\mathcal{H}_0$ is the isospin
limit one and $\mathcal{H}_1$ is the leading order breaking. 
In general for decays into final states $\Lambda f$ and $\Sigma^0 f$
we can write
\begin{align}
\bra{f\, \Sigma^0_{\mathrm{phys}}} \mathcal{H}\ket{\Lambda_{b}} &= 
       \sin\theta_m \bra{f\, \Lambda} \mathcal{H}\ket{\Lambda_b}  + 
	\cos\theta_m \bra{f\, \Sigma^0}  \mathcal{H}\ket{\Lambda_b}\,
                                                                  \approx
  \\
& ~~~~~~~~
\theta_m \bra{f\, \Lambda} \mathcal{H}_0\ket{\Lambda_b}  + 
	\bra{f\, \Sigma^0}  \mathcal{H}_1\ket{\Lambda_b}\,, \nn \\
\bra{f\, \Lambda_{\mathrm{phys}}} \mathcal{H}\ket{\Lambda_{b}} &= 
       \cos\theta_m \bra{f\, \Lambda} \mathcal{H}\ket{\Lambda_b}  - 
	\sin\theta_m \bra{f\, \Sigma^0}  \mathcal{H}\ket{\Lambda_b}\, \approx
  \bra{f\, \Lambda} \mathcal{H}_0\ket{\Lambda_b}\,,\nn
\end{align} 
where we use the isospin eigenstates $\ket{\Lambda}$ and
$\ket{\Sigma^0}$. 
It follows that we can write
\begin{align}
R \approx \theta_f \equiv \theta_m + \theta_f^{\mathrm{dyn}}\,, \qquad
\theta_f^{\mathrm{dyn}} \equiv
\frac{\bra{f\, \Sigma^0}
  \mathcal{H}_1\ket{\Lambda_b}}{\bra{f\, \Lambda}
  \mathcal{H}_0\ket{\Lambda_b}}.
\end{align}

We learn that the angle $\theta_f$ has contributions from two sources:
A universal part $\theta_m$ from wave function 
overlap, which we call \lq\lq{}static\rq\rq{} mixing, and a non-universal contribution $\theta_f^{\mathrm{dyn}}$ that we 
call \lq\lq{}dynamic\rq\rq{} mixing. We can think of
$\theta_f$ as a decay dependent \lq\lq{}effective\rq\rq{} mixing angle relevant for the decay $\Lambda_b\rightarrow \Sigma^0 f$. 
It follows 
\begin{align}
\frac{\mathcal{B}(\Lambda_b\rightarrow \Sigma^0 J/\psi)}{\mathcal{B}(\Lambda_b\rightarrow \Lambda J/\psi)} 
		  &= \frac{\mathcal{P}(\Lambda_b, \Sigma^0, J/\psi)}{\mathcal{P}(\Lambda_b, \Lambda, J/\psi)} \times \left\vert \theta_f \right\vert^2 \,. 
\end{align}
Our aim in the next section is to find $\theta_f$.

\subsection{Anatomy of $\Sigma^0$--$\Lambda$ Mixing}

We start with $\theta_m$.
Because of isospin and SU(3)$_F$ breaking effects, the physical states $\ket{\Lambda_{\mathrm{phys}}}$ and $\ket{\Sigma^0_{\mathrm{phys}}}$ deviate from their decomposition into their SU(3)$_F$ eigenstates both in the $U$-spin and in the isospin basis. As isospin is the better symmetry, we expect generically the scaling
\begin{align}
\theta_m &\sim \frac{\delta}{\varepsilon}\,. \label{eq:generic-estimate}
\end{align}
This scaling can be seen explicitly in some of the estimates of the effect.
In the quark model, the QCD part of the isospin breaking corrections comes from the strong hyperfine interaction generated by the chromomagnetic spin-spin interaction as~\cite{Franklin:1981rc}  
\begin{align}
\theta_m &= \frac{\sqrt{3}}{4} \frac{m_d-m_u}{m_s - (m_u+m_d)/2}\,, \label{eq:iso-over-SU3}
\end{align}
see also Refs.~\cite{DeRujula:1975qlm, Isgur:1979ed, Gasser:1982ap, Donoghue:1985rk, Donoghue:1989sj,Karl:1994ie}, and where constituent quark masses are used. Eq.~(\ref{eq:iso-over-SU3}) agrees with our generic estimate from group-theory considerations, Eq.~(\ref{eq:generic-estimate}).
The same analytic result, Eq.~(\ref{eq:iso-over-SU3}),~is also obtained in chiral perturbation theory~\cite{Gasser:1982ap,Gasser:1984gg}.

Within the quark model, the mixing angle can also be related to baryon masses via \cite{Franklin:1981rc, Dalitz:1964es, Gal:2015iha} 
\begin{align}
\tan\theta_m  &= \frac{\left(m_{\Sigma^0} - m_{\Sigma^+}\right) - \left(m_n - m_p\right)}{\sqrt{3} (m_{\Sigma}-m_{\Lambda}) }\,, \label{eq:mix-angle-baryon-masses-1}
\end{align}
or equally \cite{Franklin:1981rc, Gal:1967oxu, Gal:2015iha} 
\begin{align}
\tan\theta_m  &= \frac{\left( m_{\Xi^-} - m_{\Xi^0} \right) -\left( m_{\Xi^{*-}} - m_{\Xi^{*0}} \right) }{2\sqrt{3}  (m_{\Sigma}-m_{\Lambda}) }\,. \label{eq:mix-angle-baryon-masses-2}
\end{align}
In Ref.~\cite{Franklin:1981rc} Eqs.~(\ref{eq:iso-over-SU3})-(\ref{eq:mix-angle-baryon-masses-2}) have been derived within the generic \lq\lq{}independent quark model\rq\rq{}~\cite{Federman:1981nk, Rubinstein:1967zz}. Furthermore, Ref.~\cite{Franklin:1981rc} provides SU(3)-breaking corrections to Eq.~(\ref{eq:mix-angle-baryon-masses-1}) within this model.

Note that Eqs.~(\ref{eq:mix-angle-baryon-masses-1}) and (\ref{eq:mix-angle-baryon-masses-2}) automatically include also QED corrections through the measured baryon masses. Recently, lattice calculations of $\theta_m$ have become available that include QCD and QED effects \cite{Kordov:2019oer}, and which we consider as the most reliable and robust of the quoted results.

The various results for the mixing angle from the literature are summarized in Table~\ref{tab:mix-angle}. 
It turns out that the quark-model predictions agree quite well with modern lattice QCD calculations.
Note however, that the lattice result of Ref.~\cite{Kordov:2019oer} (see Table~\ref{tab:mix-angle}) demonstrates that the 
QED correction is large, contrary to the quark model expectation in Ref.~\cite{Isgur:1979ed}, and amounts to about 50\% of the total result~\cite{Kordov:2019oer}.

In the literature the mixing angle has often been assumed to be universal and employed straight forward for the prediction of decays, see Refs.~\cite{Dalitz:1964es, Karl:1994ie,Henley:1994ct, Maltman:1995qw}.
It was already pointed out in Refs.~\cite{Karl:1994ie, Karl:1999te} that the $\Sigma^0$--$\Lambda$ mixing angle can also be extracted from semileptonic $\Sigma^-\rightarrow \Lambda l^-\nu$ decays. 
The angle has also been directly related to the $\pi$--$\eta$ mixing angle \cite{Maltman:1995qw,Na:1997am}. The ratio on the right hand side of Eq.~(\ref{eq:iso-over-SU3}) can be extracted from $\eta \rightarrow 3\pi$ decays~\cite{Na:1997am} or from the comparison of $K^+\rightarrow \pi^0 e^+\nu_e$ and $K_L^0\rightarrow \pi^- e^+\nu_e$ \cite{Na:1997am}.
For pseudoscalar mesons it has been shown~\cite{Gross:1979ur, Kroll:2005sd} that the reduction of isospin violation 
from $(m_d-m_u)/(m_d+m_u) $ to the ratio in Eq.~(\ref{eq:iso-over-SU3}) is related to the Adler-Bell-Jackiw anomaly \cite{Adler:1969gk, Bell:1969ts} of QCD. 

Note that in principle also $\theta_m$ is scale dependent~\cite{Karl:1994ie}, as was 
observed for the similar case of $\pi^0$--$\eta$ mixing in Ref.~\cite{Maltman:1992my}. 
Furthermore, $\theta_m$ has an electromagnetic component. 
Depending on the relevant scale of the process in principle the QED correction can be large. 
We see from the lattice results in Table~\ref{tab:mix-angle} that this is the case for $\theta_m$. 
Very generally, at high energy scales electromagnetic interactions will dominate over QCD ones~\cite{Berlad:1973vt}. 

\subsection{The Dynamic Contribution}

The dynamical contributions to isospin breaking can be parametrized as
part of the isospin- and SU(3)$_F$-breaking expansion, see
Sec.~\ref{sec:su3breaking}. Explicitly we found
\begin{align}
\theta_{J/\psi}^{\mathrm{dyn}} \equiv
\frac{\bra{J/\psi\, \Sigma^0}
  \mathcal{H}_1\ket{\Lambda_b}}{\bra{J/\psi\, \Lambda}
  \mathcal{H}_0\ket{\Lambda_b}} = \delta \times 
\left[\sqrt{1\over 5}{B^{15} \over A^3_c} - \sqrt{1\over 2}{B^{\bar 6} \over A^3_c}\right].
\end{align}
We expect that $B^{15} \sim B^{\bar 6} \sim A^3_c$.
The important result is that these effects
are order $\delta$.
Taking everything into account, very schematically we expect therefore the power counting
\begin{align}
\theta_f \sim \left(\frac{\delta}{\varepsilon}\right)_m + \delta_f \sim  \theta_m \left[1 + \mathcal{O}(\varepsilon_f) \right]\,. \label{eq:theta-power-counting}
\end{align}
where $\delta_f$ and $\varepsilon_f$ refer to isospin and $SU(3)$ breaking
parameters that depend on $f$.

\subsection{Prediction for $\mathcal{B}(\Lambda_b\rightarrow \Sigma^0 J/\psi)$ \label{sec:dynamic}} 

We see from the power counting in Eq.~(\ref{eq:theta-power-counting}) that the static component $\theta_m$ dominates, as it is relatively enhanced by the inverse of the size of SU(3)$_F$ breaking.
Employing this assumption we obtain for
\begin{align}
\theta_f\sim \theta_m \sim 1^{\circ} 
\end{align}
the prediction
\begin{align}
\left\vert\frac{\mathcal{A}(\Lambda_b\rightarrow \Sigma^0 J/\psi)}{\mathcal{A}(\Lambda_b\rightarrow \Lambda J/\psi)}\right\vert &=  \left\vert \theta_f \right\vert \sim 0.02 \,. \label{eq:sigma-prediction} 
\end{align}  
A confirmation of our prediction would imply the approximate universality of the
$\Sigma^0$--$\Lambda$ mixing angle in $b$-baryon decays. In that case we would expect that likewise 
\begin{align}
\left\vert\frac{\mathcal{A}(\Lambda_b\rightarrow \Sigma^0 J/\psi)}{\mathcal{A}(\Lambda_b\rightarrow \Lambda J/\psi)}\right\vert &=  
\left\vert\frac{\mathcal{A}(\Lambda_b\rightarrow \Sigma^0\gamma) }{\mathcal{A}(\Lambda_b\rightarrow \Lambda\gamma)} \right\vert=
\left\vert\frac{\mathcal{A}(\Lambda_b\rightarrow \Sigma^0 l^+l^-) }{\mathcal{A}(\Lambda_b\rightarrow \Lambda l^+l^-)} \right\vert=
\left\vert\frac{\mathcal{A}(\Sigma_b^0 \rightarrow \Lambda J/\psi)}{\mathcal{A}(\Sigma_b^0 \rightarrow \Sigma^0  J/\psi)} \right\vert
\sim 0.02\,,
\end{align} 
up to SU(3)$_F$ breaking.
Note that $\Lambda_b$ and $\Sigma_b$ are not in the same SU(3)$_F$
multiplet, so that there is no relation between their reduced matrix
elements. 

The above predictions are based on the assumption that the dynamic
contribution is smaller by a factor of the order of the SU(3) breaking. In practice,
these effects may be large enough to be probed experimentally. Thus, we
can hope that precise measurements of these ratios will be able to test these assumptions.

\section{Comparison with recent data \label{sec:data}}

We move to compare the general results of Sections~\ref{sec:group-theory} and~\ref{sec:mixing} to the recent LHCb data for the case $S=J/\psi$~\cite{Aaij:2019hyu}. Particularly relevant to the experimental findings is the sum rule Eq.~(\ref{eq:key-sum-rule}) which we rephrase as 
\begin{align}
\left\vert \frac{\mathcal{A}(\Xi_b^0\rightarrow \Lambda J/\psi)}{\mathcal{A}(\Xi_b^0\rightarrow \Xi^0 J/\psi)}\right\vert = \frac{1}{\sqrt{6}} \left(1+\mathcal{O}(\varepsilon) \right)\left\vert\frac{\lambda_{cd}}{\lambda_{cs} }\right\vert  \approx 0.41\, \left\vert\frac{\lambda_{cd}}{\lambda_{cs} }\right\vert  \,, \label{eq:SU3relation} 
\end{align} 
where in the last step we only wrote the central value. The error is expected to be roughly of order $\varepsilon\sim 20\%$.
The estimate in Eq.~(\ref{eq:SU3relation})
agrees very well with the recent measurement~\cite{Aaij:2019hyu}
\begin{align}
\left\vert \frac{\mathcal{A}(\Xi_b^0\rightarrow \Lambda J/\psi)}{\mathcal{A}(\Xi_b^0\rightarrow \Xi^0 J/\psi)} \right\vert = \left(0.44\pm 0.06\pm 0.02\right)  \left\vert\frac{\lambda_{cd}}{\lambda_{cs} }\right\vert \,. \label{eq:SU3relmeasurement}
\end{align}
This suggests that the assumptions made in Sec.~\ref{sec:group-theory} are justified. However, from the SU(3)$_F$-breaking contributions which we calculated in Sec.~\ref{sec:su3breaking}, we expect generically an order 20\% correction to Eq.~(\ref{eq:SU3relation}). The measurement Eq.~(\ref{eq:SU3relmeasurement}) is not yet precise enough to probe and learn about the size of these corrections. However, SU(3)$_F$ breaking seems also not to be enhanced beyond the generic 20\%. 

The only other theory result for the ratio Eq.~(\ref{eq:SU3relmeasurement}) that we are aware of in the literature can be obtained from the branching ratios
provided in Ref.~\cite{Gutsche:2018utw}, where a covariant confined quark model has been employed. From the branching ratios given therein we extract the
central value 
\begin{align} 
\left\vert \frac{\mathcal{A}(\Xi_b^0\rightarrow \Lambda J/\psi)}{\mathcal{A}(\Xi_b^0\rightarrow \Xi^0 J/\psi)} \right\vert 
	\sim 0.34 \,\frac{\lambda_{cd} }{\lambda_{cs}}\,, 
\end{align}
where an error of $\sim 20\%$ is quoted in Ref.~\cite{Gutsche:2018utw} for the branching ratios. 
This estimate is also in agreement with the data, Eq.~(\ref{eq:SU3relmeasurement}) 
(see for details in Ref.~\cite{Gutsche:2018utw}).

Finally, our prediction Eq.~(\ref{eq:sigma-prediction})
\begin{align}
\left\vert\frac{\mathcal{A}(\Lambda_b\rightarrow \Sigma^0 J/\psi)}{\mathcal{A}(\Lambda_b\rightarrow \Lambda J/\psi)}\right\vert &=  \left\vert \theta_f \right\vert \sim 0.02 \,, 
\end{align}  
is only about a factor two below the bound provided in Ref.~\cite{Aaij:2019hyu}, 
\begin{align}
\left\vert\frac{\mathcal{A}(\Lambda_b\rightarrow \Sigma^0 J/\psi)}{\mathcal{A}(\Lambda_b\rightarrow \Lambda J/\psi)}\right\vert 
	< 1/20.9 = 0.048  \quad \text{at $95\%$ CL.} \label{eq:sigma-bound}
\end{align}
A deviation from Eq.~(\ref{eq:sigma-prediction}) would indicate the observation of a non-universal contribution to the effective mixing angle, i.e.~an enhancement of isospin violation in the dynamical contribution~$\theta^{\rm dyn}_{J/\Psi}$.  
It seems that a first observation of isospin violation in $\Lambda_b$ decays is feasible for LHCb in the near future.

\section{Conclusions \label{sec:conclusions}}

We perform a comprehensive SU(3)$_F$ analysis of two-body $b\rightarrow c\bar{c}s(d)$ decays of the $b$-baryon antitriplet to baryons of the light octet and an SU(3)$_F$ singlet, including a discussion of isospin and SU(3)$_F$ breaking effects as well as $\Sigma^0$--$\Lambda$ mixing.
Our formalism allows us to interpret recent results for the case $S=J/\psi$ by LHCb, which do not yet show signs of isospin violation or SU(3)$_F$ breaking. 
We point out several sum rules which can be tested in the future and give a prediction for the ratios   
$\vert \mathcal{A}(\Lambda_b\rightarrow \Sigma^0 J/\psi)\vert / \vert \mathcal{A}(\Lambda_b\rightarrow \Lambda J/\psi)\vert \sim 0.02$ and
$\left\vert \mathcal{A}(\Xi_b^0\rightarrow \Lambda J/\psi)/ \mathcal{A}(\Xi_b^0\rightarrow \Xi^0 J/\psi)\right\vert\approx 1/\sqrt{6}\, \left\vert V_{cb}^* V_{cd} / (V_{cb}^* V_{cs})\right\vert$.
More measurements are needed in order to probe isospin and SU(3)$_F$ breaking corrections to these and many more relations that we laid out in this work.

\begin{acknowledgments}
We thank Sheldon Stone for discussions.
The work of YG is supported in part by the NSF grant PHY1316222.
SS is supported by a DFG Forschungs\-stipendium under contract no. SCHA 2125/1-1.
\end{acknowledgments}

\begin{table}[b]
\begin{center}
\begin{tabular}{c|c|c|c|c|c}
\hline \hline
Particle          & Quark Content 
		  & SU(3)$_F$ State
		  & Isospin
		  & $U$-spin
		  & Hadron Mass [MeV] 
		  \\\hline 
%
 $u$		  & $u$ 			        
	          & $\ket{\mathbf{3}}_{\frac{1}{2},\frac{1}{2}, \frac{1}{3}}$
		  & $\ket{\frac{1}{2},\frac{1}{2}}_I$   
		  & $\ket{0,0}_U$
		  & n/a  \\  
 $d$		  & $d$   				
	          & $\ket{\mathbf{3}}_{\frac{1}{2},-\frac{1}{2},\frac{1}{3}}$
		  & $\ket{\frac{1}{2},-\frac{1}{2}}_I$
		  & $\ket{\frac{1}{2},\frac{1}{2} }_U$ 
		  & n/a  \\                     	
 $s$		  & $s$   				
	          & $\ket{\mathbf{3}}_{0,0,-\frac{2}{3}}$ 
		  & $\ket{0,0}_I$
		  & $\ket{\frac{1}{2},-\frac{1}{2}}_U$
		  & n/a \\\hline             
 $\ket{\Lambda_b}$& $udb$	 			
	          & $\ket{\mathbf{\bar{3}}}_{0,0,\frac{2}{3}}$ 
		  &  $\ket{0,0}_I$                    
		  &  $\ket{\frac{1}{2},\frac{1}{2}}_U$
		  &  $5619.60\pm 0.17$ \\ 	                                
$\ket{\Xi_b^-}$	  & $dsb$	 	
		  & $\ket{\mathbf{\bar{3}}}_{\frac{1}{2},-\frac{1}{2},-\frac{1}{3}}$
		  & $\ket{\frac{1}{2},-\frac{1}{2}}_I$  
		  & $\ket{0,0}_U$                      
		  & $5797.0\pm 0.9$   \\ 	                
$\ket{\Xi_b^0}$   & $usb$ 	 
		  & $\ket{\mathbf{\bar{3}}}_{\frac{1}{2},\frac{1}{2},-\frac{1}{3}}$ 
		  & $\ket{\frac{1}{2},\frac{1}{2}}_I$  
		  & $\ket{\frac{1}{2}, -\frac{1}{2}}_U$
		  & $5791.9\pm 0.5$  \\\hline 	         	 
 $\ket{\Lambda}$  & $uds$	 				 
	          & $\ket{\mathbf{8}}_{0,0,0}$ 
		  & $\ket{0,0}_I$                               
		  & $\frac{\sqrt{3}}{2}\ket{1,0}_U -\frac{1}{2}\ket{0,0}_U$ 
		  & $1115.683\pm 0.006$   \\ 	                         
 $\ket{\Sigma^0}$ & $uds$	 				 
	          & $\ket{\mathbf{8}}_{1,0,0}$ 
		  & $\ket{1,0}_I$                 
		  & $\frac{1}{2} \ket{1,0}_U +  \frac{\sqrt{3}}{2} \ket{0,0}_U$   
		  &  $1192.642\pm 0.024$    \\ 	                 	 
$\ket{\Sigma^-}$ & $dds$	 
	          & $\ket{\mathbf{8}}_{1,-1,0}$  
		  & $\ket{1,-1}_I$         
		  & $\ket{\frac{1}{2},\frac{1}{2}}_U$     
		  & $1197.449\pm 0.0030$        \\ 	                 
$\ket{\Sigma^+}$ & $uus$	 
	          & $\ket{\mathbf{8}}_{1, 1, 0}$ 
		  & $\ket{1,1 }_I$ 
		  & $\ket{\frac{1}{2},-\frac{1}{2}}_U$ 
		  & $1189.37\pm 0.07$       \\ 	                 
  $\ket{\Xi^0}$	  & $uss$	 
	          & $\ket{\mathbf{8}}_{\frac{1}{2},\frac{1}{2},-1}$ 
		  & $\ket{\frac{1}{2},\frac{1}{2}}_I   $                 
		  & $\ket{1,-1}_U$                                     
		  &  $1314.86\pm 0.20$     \\ 	                 
  $\ket{\Xi^-}$	  & $dss$	 
	          & $\ket{\mathbf{8}}_{\frac{1}{2},-\frac{1}{2},-1}$ 
		  & $\ket{\frac{1}{2}, -\frac{1}{2}}_I$   
		  & $\ket{\frac{1}{2},-\frac{1}{2}}_U$    
		  &  $1321.71 \pm 0.07$       \\        
  $\ket{n}$ 	   & $udd$	 			 
	           & $\ket{\mathbf{8}}_{\frac{1}{2},-\frac{1}{2},1}$ 
		   & $\ket{\frac{1}{2},-\frac{1}{2}}_I$         
		   & $\ket{1,1}_U$                                      
		   & $939.565413\pm 0.000006$    \\ 	                 
  $\ket{p}$ 	   & $uud$	 			 
	           & $\ket{\mathbf{8}}_{\frac{1}{2},\frac{1}{2},1}$  
		   & $\ket{\frac{1}{2},\frac{1}{2}}_I$   
		   & $\ket{\frac{1}{2},\frac{1}{2}}_U$    
		   & $938.2720813\pm 0.0000058$   \\\hline 	                 
 $\ket{J/\psi}$	  & $c\bar{c}$	 			 
	          & $\ket{\mathbf{1}}_{0,0,0}$ 
		  & $\ket{0,0}_I$                       
		  & $\ket{0,0}_U$                       
		  & $3096.900\pm 0.006$      \\\hline\hline 	                 
\end{tabular}
\caption{SU(3)$_F$, isospin and $U$-spin wave functions~\cite{deSwart:1963pdg, Ali:2012pn, Tanabashi:2018oca, Roberts:2007ni, Greiner:1989eu} and masses~\cite{Tanabashi:2018oca}. For the indices of the SU(3)$_F$ states we use the convention $\ket{\mu}_{I, I_3, Y}$. 
The lifetimes of the members of the heavy baryon triplet are 
$\tau_{\Lambda_b}/\tau_{B^0}  = 0.964\pm 0.007$\,, where  
$\tau_{B^0} = (1519\pm 4)\times 10^{-15} s$\,,
$\tau_{\Xi_b^0} = (1.480\pm 0.030)\times 10^{-12}\, s$\,,
$\tau_{\Xi_b^-} = (1.572\pm 0.040)\cdot 10^{-12}\, s$\,. 
\cite{Tanabashi:2018oca}. 
Note that the exact form of the $\Sigma^0$--$\Lambda$ mixing is phase convention dependent~\cite{Greiner:1989eu}.
Our convention agrees with Refs.~\cite{Greiner:1989eu, Rosner:1981yh}. 
Another convention can be found e.g. in Ref.~\cite{Novozhilov:1975yt} in the form of 
$\ket{\Lambda} =\frac{\sqrt{3}}{2}\ket{1,0}_U - \frac{1}{2} \ket{0,0}_U$ and 
$\ket{\Sigma^0} = -\frac{1}{2} \ket{1,0}_U -\frac{\sqrt{3}}{2} \ket{0,0}_U$\,. 
\label{tab:particles}}
\end{center}
\end{table}

\begin{table}[t]
\begin{center}
\begin{tabular}{c|c|c|c|c}
\hline \hline

Decay ampl. $\mathcal{A}$ 
		& $A_c^3$
		& $A_u^3$ 
		& $A_u^{\overline{6}}$ 
		& $A_u^{15}$ \\\hline 

\multicolumn{5}{c}{$b\rightarrow s$} \\\hline 

$\mathcal{A}(\Lambda_b \rightarrow  \Lambda  S)$ & $\sqrt{\frac{2}{3}}  \lambda_{cs}$  & $\sqrt{\frac{2}{3}} \lambda_{us}$  & 0  & $\sqrt{\frac{6}{5}} \lambda_{us}$  
\\ 

$\mathcal{A}(\Lambda_b \rightarrow  \Sigma^0 S)$ &  $0$  & $0$ &$ -\sqrt{\frac{2}{3}} \lambda_{us}$ & $2\sqrt{\frac{2}{5}} \lambda_{us}$  
\\ 

$\mathcal{A}(\Xi_b^0   \rightarrow  \Xi^0    S)$ & $\lambda_{cs}$  &  $\lambda_{us}$  &  $\sqrt{\frac{1}{3}} \lambda_{us}$ & $\sqrt{\frac{1}{5}} \lambda_{us}$  
\\ 

$\mathcal{A}(\Xi_b^-   \rightarrow  \Xi^-    S)$ & $\lambda_{cs}$  & $\lambda_{us}$ & $-\sqrt{\frac{1}{3}} \lambda_{us}$   &  $-\frac{3}{\sqrt{5}} \lambda_{us}$  
\\\hline 

\multicolumn{5}{c}{$b\rightarrow d$} \\\hline 

$\mathcal{A}(\Xi_b^0 \rightarrow \Lambda S)$  & $-\sqrt{\frac{1}{6}}\lambda_{cd}$  &  $-\sqrt{\frac{1}{6}}\lambda_{ud}$  & $-\frac{1}{\sqrt{2}}  \lambda_{ud}$  & $\sqrt{\frac{3}{10}} \lambda_{ud}$  \\ 
                                                                
$\mathcal{A}(\Xi_b^0 \rightarrow \Sigma^0 S)$ & $\frac{1}{\sqrt{2}}\lambda_{cd}$   & $\frac{1}{\sqrt{2}}\lambda_{ud}$   & $-\frac{1}{\sqrt{6}} \lambda_{ud}$    & $\sqrt{\frac{5}{2}} \lambda_{ud}$   \\ 
                                                                
$\mathcal{A}(\Lambda_b \rightarrow n S)$      & $\lambda_{cd}$ & $\lambda_{ud}$ & $\frac{1}{\sqrt{3}} \lambda_{ud}$  & $\frac{1}{\sqrt{5}} \lambda_{ud}$  \\ 
                                                                
$\mathcal{A}(\Xi_b^- \rightarrow \Sigma^- S)$ & $\lambda_{cd}$   & $\lambda_{ud}$ & $-\frac{1}{\sqrt{3}}\lambda_{ud} $   &  $ -\frac{3}{\sqrt{5}} \lambda_{ud}$  \\ 

\hline\hline
\end{tabular}
\caption{SU(3)$_F$-limit decomposition. \label{tab:SU3}}
\end{center}
\end{table}

\begin{table}[t]
\begin{center}
\begin{tabular}{c|c|c|c|c|c|c}
\hline \hline
\multicolumn{7}{c}{$b\rightarrow s$} \\\hline
Decay Ampl.  $\mathcal{A}$
                & $\bra{0}0\ket{0}^c$
                & $\bra{\frac{1}{2}} 0 \ket{\frac{1}{2}}^c$
                & $\bra{0}0\ket{0}^u$
                & $\bra{1}1\ket{0}^u$
                & $\bra{\frac{1}{2}} 1 \ket{\frac{1}{2}}^u$
		& $\bra{\frac{1}{2}} 0 \ket{\frac{1}{2}}^u$
\\\hline
 $\mathcal{A}(\Lambda_b \rightarrow  \Lambda  S)$ & $\lambda_{cs}$ & 0 & $\lambda_{us}$ & 0 & 0 & 0\\

 $\mathcal{A}(\Lambda_b \rightarrow  \Sigma^0 S)$ & 0 & 0 & 0 & $\lambda_{us}$ & 0 & 0\\

 $\mathcal{A}(\Xi_b^0   \rightarrow  \Xi^0    S)$ & 0 & $\lambda_{cs}$ & 0& 0& $-\sqrt{\frac{1}{3}} \lambda_{us}$ & $\lambda_{us}$ \\

 $\mathcal{A}(\Xi_b^-   \rightarrow  \Xi^-    S)$ &  0 & $\lambda_{cs}$ & 0 & 0 & $\sqrt{\frac{1}{3}}\lambda_{us}$ & $\lambda_{us}$\\
\hline\hline
\end{tabular}
\caption{Isospin decomposition for $b\rightarrow s$ transitions. \label{tab:isospin1}}
\end{center}
\end{table}

\begin{table}[t]
\begin{center}
\begin{tabular}{c|c|c|c|c|c|c|c}
\hline \hline

\multicolumn{8}{c}{$b\rightarrow d$} \\\hline

Decay ampl. $\mathcal{A}$
                & $\bra{0} \frac{1}{2} \ket{\frac{1}{2}}^c$ & $\bra{0} \frac{1}{2} \ket{\frac{1}{2}}^u$
                & $\bra{1}\frac{1}{2} \ket{\frac{1}{2}}^c$ &  $\bra{1} \frac{1}{2} \ket{\frac{1}{2}}^u$
                & $\bra{\frac{1}{2}}\frac{1}{2} \ket{0}^c$ &  $\bra{\frac{1}{2}} \frac{1}{2} \ket{0}^u$
                & $\bra{1} \frac{3}{2} \ket{\frac{1}{2}}^u$ \\\hline

$\mathcal{A}(\Xi_b^0 \rightarrow \Lambda S)$  & $-\frac{1}{\sqrt{2}}\lambda_{cd}$  &  $-\frac{1}{\sqrt{2}}\lambda_{ud}$ & 0  & 0 & 0 & 0 & 0  \\

$\mathcal{A}(\Xi_b^0 \rightarrow \Sigma^0 S)$ & 0 & 0& $\frac{1}{\sqrt{2}}\lambda_{cd}$  & $\frac{1}{\sqrt{2}}\lambda_{ud} $ & 0  & 0 & $-\frac{1}{\sqrt{2}}  \lambda_{ud}$ \\

$\mathcal{A}(\Lambda_b \rightarrow n S)$      & 0 &  0 & 0  & 0 & $\lambda_{cd}$ & $\lambda_{ud}$ & 0\\

$\mathcal{A}(\Xi_b^- \rightarrow \Sigma^- S)$ & 0 & 0 & $\lambda_{cd}$ & $\lambda_{ud}$ & 0 & 0 & $\frac{1}{2} \lambda_{ud}$  \\
\hline\hline
\end{tabular}
\caption{Isospin decomposition for $b\rightarrow d$ transitions. \label{tab:isospin2}}
\end{center}
\end{table}

\begin{table}[t]
\begin{center}
\begin{tabular}{c|c|c|c|c|c|c}
\hline \hline

Decay ampl. $\mathcal{A}$ 
                & $\bra{0}\frac{1}{2}\ket{\frac{1}{2}}^c$   & $\bra{0}\frac{1}{2}\ket{\frac{1}{2}}^u$
                & $\bra{1} \frac{1}{2} \ket{\frac{1}{2}}^c$ & $\bra{1} \frac{1}{2} \ket{\frac{1}{2}}^u$
                & $\bra{\frac{1}{2}} \frac{1}{2} \ket{0}^c$ & $\bra{\frac{1}{2}} \frac{1}{2} \ket{0}^u$  \\\hline 

\multicolumn{7}{c}{$b\rightarrow s$} \\\hline 

$\mathcal{A}(\Lambda_b \rightarrow  \Lambda  S)$ & $\frac{1}{2\sqrt{2}}\lambda_{cs}$
						 & $\frac{1}{2\sqrt{2}}\lambda_{us}$
						 & $\frac{\sqrt{3}}{2\sqrt{2}}\lambda_{cs}$
						 & $\frac{\sqrt{3}}{2\sqrt{2}}\lambda_{us}$
						 & $0$ 
						 & $0$ \\ 

$\mathcal{A}(\Lambda_b \rightarrow  \Sigma^0 S)$ & $-\frac{\sqrt{3}}{2\sqrt{2}}\lambda_{cs}$
						 & $-\frac{\sqrt{3}}{2\sqrt{2}}\lambda_{us}$
						 & $\frac{1}{2\sqrt{2}} \lambda_{cs}$ 
						 & $\frac{1}{2\sqrt{2}} \lambda_{us}$ 
						 & $0$  
						 & $0$  \\   

$\mathcal{A}(\Xi_b^0   \rightarrow  \Xi^0    S)$ & $0$
						 & $0$
						 & $\lambda_{cs}$ 
						 & $\lambda_{us}$ 
						 & $0$
						 & $0$	\\  

$\mathcal{A}(\Xi_b^-   \rightarrow  \Xi^-    S)$ & $0$
						 & $0$
						 & $0$ 
						 & $0$ 
						 & $\lambda_{cs}$ 
						 & $\lambda_{us}$  \\\hline 

\multicolumn{7}{c}{$b\rightarrow d$} \\\hline 

$\mathcal{A}(\Xi_b^0 \rightarrow \Lambda S)$  & $-\frac{1}{2\sqrt{2}}\lambda_{cd}$
					      & $-\frac{1}{2\sqrt{2}}\lambda_{ud}$
					      & $\frac{\sqrt{3}}{2\sqrt{2}}\lambda_{cd}$ 
					      & $\frac{\sqrt{3}}{2\sqrt{2}}\lambda_{ud}$ 
					      & 0 
				              & 0 \\ 

$\mathcal{A}(\Xi_b^0 \rightarrow \Sigma^0 S)$ & $\frac{\sqrt{3}}{2\sqrt{2}}\lambda_{cd}$  
					      & $\frac{\sqrt{3}}{2\sqrt{2}}\lambda_{ud}$  
					      & $\frac{1}{2\sqrt{2}}\lambda_{cd}$  
					      & $\frac{1}{2\sqrt{2}}\lambda_{ud}$  
					      & 0  
				  	      & 0 \\   

$\mathcal{A}(\Lambda_b \rightarrow n S)$      & 0 
					      & 0
					      & $\lambda_{cd}$ 
					      & $\lambda_{ud}$
					      & 0 
					      & 0 \\ 

$\mathcal{A}(\Xi_b^- \rightarrow \Sigma^- S)$ & 0
					      & 0
					      & 0 
					      & 0
					      & $\lambda_{cd}$ 
				 	      & $\lambda_{ud}$	\\ 
\hline\hline
\end{tabular}
\caption{$U$-spin decomposition. \label{tab:uspin}}
\end{center}
\end{table}

\begin{table}[t]
\begin{center}
\begin{tabular}{c|c|c|c|c|c|c}
\hline \hline

Decay ampl. $\mathcal{A}$ 
                & $\bra{0}\frac{1}{2}\ket{\frac{1}{2}}^c$   & $\bra{0}\frac{1}{2}\ket{\frac{1}{2}}^u$
                & $\bra{1} \frac{1}{2} \ket{\frac{1}{2}}^c$ & $\bra{1} \frac{1}{2} \ket{\frac{1}{2}}^u$
                & $\bra{\frac{1}{2}} \frac{1}{2} \ket{0}^c$ & $\bra{\frac{1}{2}} \frac{1}{2} \ket{0}^u$  \\\hline 

\multicolumn{7}{c}{$b\rightarrow s$} \\\hline 

$\mathcal{A}(\Lambda_b \rightarrow  X  S)$ & $-\frac{1}{\sqrt{2}} \lambda_{cs}$
					   & $-\frac{1}{\sqrt{2}} \lambda_{us}$
					   & 0 
  					   & 0 
					   & $0$ 
					   & $0$ \\ 

$\mathcal{A}(\Lambda_b \rightarrow  Y S)$ & 0  
					  & 0  
					  & $\frac{1}{\sqrt{2}} \lambda_{cs}$ 
					  & $\frac{1}{\sqrt{2}} \lambda_{us}$ 
					  & $0$  
					  & $0$  	\\\hline 

\multicolumn{7}{c}{$b\rightarrow d$} \\\hline 

$\mathcal{A}(\Xi_b^0 \rightarrow X S)$     & $\frac{1}{\sqrt{2}} \lambda_{cd}$
					   & $\frac{1}{\sqrt{2}} \lambda_{ud}$
				           & 0  
					   & 0
					   & 0
					   & 0 \\ 

$\mathcal{A}(\Xi_b^0 \rightarrow Y S)$  & 0  
					& 0  
					& $\frac{1}{\sqrt{2}} \lambda_{cd}$  
					& $\frac{1}{\sqrt{2}} \lambda_{ud}$  
					& 0  
					& 0 \\   

\hline\hline
\end{tabular}
\caption{(Unpractical) $U$-spin decomposition for the $U$-spin eigenstates $\ket{X}$ and $\ket{Y}$, see Eq.~(\ref{eq:hypothetical}) and discussion in the text.  \label{tab:uspin-hypo}}
\end{center}
\end{table}

\begin{table}[t]
\begin{center}
\begin{tabular}{c|c|c}
\hline \hline
Method           			& Mixing Angle $\theta_m$ $[^\circ{}] $ & Ref. \\\hline 

Quark model: Relation to Baryon masses    & $0.86\pm 0.06$   &    \cite{Dalitz:1964es,Gal:2015iha}  \\

Quark model: Hyperfine splitting + EM interactions  & $\simeq 0.57$	 & \cite{Isgur:1979ed}     \\

Lattice QCD+QED A			& $1.00\pm 0.32$	& \cite{Kordov:2019oer}    \\

Lattice QCD+QED B			& $0.96\pm 0.31$	& \cite{Kordov:2019oer}   \\

Lattice QCD without QED		        & $0.55\pm 0.03$	& \cite{Kordov:2019oer} \\\hline\hline
\end{tabular}
\caption{Results for the $\Sigma^0$--$\Lambda$ mixing angle. Note that we adjusted the sign conventions for the results to match always the one of Ref.~\cite{Dalitz:1964es}, see also the corresponding comment in Ref.~\cite{Kordov:2019oer}. 
For older lattice results for the \lq\lq{}QCD only\rq\rq{} scenario see Refs.~\cite{Horsley:2014koa, Horsley:2015lha}.
Note that with alternate quark mass input taken from Ref.~\cite{Aoki:2019cca} the result for the \lq\lq{}Lattice QCD without QED\rq\rq{} scenario is changed to $\theta_m = 0.65\pm 0.03$~\cite{Kordov:2019oer}. 
\label{tab:mix-angle}}
\end{center}
\end{table}

\begin{table}[t]
\begin{center}
\begin{tabular}{c|c||c|c|c}
\hline \hline

Decay ampl. $\mathcal{A}$ 
		& $A_c^3$
		& $B^{3}$ 
		& $B^{15}$ 
		& $B^{\overline{6}}$ 
		 \\\hline 

\multicolumn{5}{c}{$b\rightarrow s$} \\\hline 

$\mathcal{A}(\Lambda_b \rightarrow  \Lambda  S)/\lambda_{cs}$ & $\sqrt{\frac{2}{3}}$    
& $\frac{1}{2}\sqrt{\frac{1}{3}}\,\varepsilon$   & $\sqrt{\frac{3}{10}}\,\varepsilon$  & $0$   \\ 

$\mathcal{A}(\Lambda_b \rightarrow  \Sigma^0 S) /\lambda_{cs}$ &  $0$  
& $0$  & $\sqrt{\frac{2}{15}}\,\delta$ & $-\sqrt{\frac{1}{3}}\,\delta$      \\ 

$\mathcal{A}(\Xi_b^0   \rightarrow  \Xi^0    S) /\lambda_{cs}$ & $1$  
& $\frac{1}{2}\,\varepsilon$  &  $\sqrt{\frac{1}{15}}\,\delta - \frac{1}{2\sqrt{5}}\,\varepsilon$  & $\sqrt{\frac{1}{6}}\,\delta$  \\ 

$\mathcal{A}(\Xi_b^-   \rightarrow  \Xi^-    S) /\lambda_{cs}$ & $1$  
& $\frac{1}{2}\,\varepsilon$ &  $-\sqrt{\frac{1}{15}}\,\delta - \frac{1}{2\sqrt{5}}\,\varepsilon$  & $-\sqrt{\frac{1}{6}}\,\delta$ \\\hline 

\multicolumn{5}{c}{$b\rightarrow d$} \\\hline 

$\mathcal{A}(\Xi_b^0 \rightarrow \Lambda S)/\lambda_{cd}$  & $-\frac{1}{\sqrt{6}}$  
		& $ -\frac{1}{4\sqrt{2}} \delta +  \frac{1}{4\sqrt{6}} \varepsilon$ 
		& $-\frac{1}{4\sqrt{10}} \delta +  \frac{3}{4} \sqrt{\frac{3}{10}} \varepsilon$    
		& $-\frac{1}{4} \delta - \frac{\sqrt{3}}{4} \varepsilon$             \\ 

$\mathcal{A}(\Xi_b^0 \rightarrow \Sigma^0 S)/\lambda_{cd}$ & $\frac{1}{\sqrt{2}}$   
		& $\frac{1}{4}\sqrt{\frac{3}{2}} \delta  - \frac{1}{4\sqrt{2}} \varepsilon  $   
		&  $\frac{11}{4\sqrt{30}} \delta  - \frac{1}{4\sqrt{10}}\varepsilon $          
		& $-\frac{1}{4\sqrt{3}} \delta  - \frac{1}{4}  \varepsilon $    \\   

$\mathcal{A}(\Lambda_b \rightarrow n S)/\lambda_{cd}$      & $1$  
		& $\frac{\sqrt{3}}{4} \delta   - \frac{1}{4} \varepsilon$  
		& $-\frac{1}{4\sqrt{15}}\delta  + \frac{3}{4\sqrt{5}}\varepsilon$      
		& $ \frac{1}{2\sqrt{6}} \delta +\frac{1}{2\sqrt{2}} \varepsilon$  \\ 

$\mathcal{A}(\Xi_b^- \rightarrow \Sigma^- S)/\lambda_{cd}$ & $1$  
		& $\frac{\sqrt{3}}{4} \delta - \frac{1}{4} \varepsilon $   
		& $ -\frac{1}{4}\sqrt{\frac{5}{3}} \delta - \frac{1}{4\sqrt{5}} \varepsilon$      
		& $ - \frac{1}{2\sqrt{6}} \delta -\frac{1}{2\sqrt{2}} \varepsilon$       \\\hline\hline
\end{tabular}
\caption{CKM-leading SU(3)$_F$ decomposition including isospin- and SU(3)$_F$-breaking. \label{tab:SU3breaking}}
\end{center}
\end{table}

\clearpage

\bibliography{draft.bib}
\bibliographystyle{apsrev4-1}

\end{document}